\documentclass[a4paper,11pt]{article}
\usepackage{jheppub} 
\usepackage{lineno}
\usepackage[dvipsnames]{xcolor}

\usepackage{caption}
\usepackage{subcaption}
\usepackage{mathtools}

\usepackage{slashed}
\usepackage{cleveref}
\usepackage[normalem]{ulem}
\usepackage{multirow}
\usepackage{soul}

\usepackage{tablefootnote}

\newcommand{\dd}{\mathrm{d}}
\newcommand{\sigmaV}{\left\langle\sigma v \right \rangle }

\title{\boldmath Dark matter phase-in: producing feebly-interacting particles after a first-order phase transition}

\preprint{P3H-25-028, TTP25-013}

\author[1,2]{Cristina Benso,}
\author[1,2]{Felix~Kahlhoefer}
\author[2]{and Henda Mansour}
\affiliation[1]{Institut für Astroteilchen Physik, Karlsruher Institut für Technologie (KIT), Hermann-von-\\Helmholtz-Platz 1, 76344 Eggenstein-Leopoldshafen, Germany}
\affiliation[2]{Institute for Theoretical Particle Physics (TTP), Karlsruhe Institute of Technology (KIT), \\D-76131 Karlsruhe, Germany}

\emailAdd{cristina.benso@kit.edu}
\emailAdd{kahlhoefer@kit.edu}
\emailAdd{henda.mansour@kit.edu}

{\abstract{
The freeze-in mechanism describes the out-of-equilibrium production of dark matter (DM) particles via feeble couplings or non-renormalisable interactions with large suppression scales. In the latter case, predictions suffer from a strong sensitivity to the initial conditions of the universe, such as the details of reheating. In this work, we investigate how this sensitivity is altered in the presence of a cosmological first-order phase transition. We show that freeze-in via non-renormalisable interactions is not always dominated by the highest temperatures of the Standard Model (SM) thermal bath, but instead may be governed by the period immediately after the phase transition, during which the decaying scalar field transfers its energy density to the SM radiation. We refer to this alternative production regime as \textit{DM phase-in}. Using numerical and approximate analytical solutions of the relevant Boltzmann equations, we determine the conditions that under which phase-in or conventional freeze-in production dominates the final DM abundance in terms of the type of interaction between the DM and SM particles, the amount of supercooling before and the evolution of the scalar field after the phase transition. In the phase-in regime, the DM abundance is correlated with the peak frequency of the gravitational wave signal associated with the phase transition, opening up new observational possibilities. }


}

\begin{document}
\maketitle
\flushbottom

\section{Introduction}

All present evidence for dark matter (DM) is based exclusively on its gravitational interactions, and the extent to which DM particles have non-gravitational interactions with known matter is completely unknown~\cite{Balazs:2024uyj}. Nevertheless, many models of DM assume that such interactions exist and that they are sufficient to bring DM into equilibrium with the Standard Model (SM) thermal bath in the very early universe, after which DM particles become Boltzmann suppressed and freeze out. The interaction strength required for this process to reproduce the observed DM abundance leads to testable predictions for laboratory experiments, which have so far not been able to confirm the freeze-out picture.

In light of these null results, an alternative DM production mechanism has gained traction in recent years, in which the non-gravitational interactions between DM and SM particles are too weak to ever bring the different species into equilibrium. In this so-called freeze-in mechanism~\cite{Hall:2009bx,Chu:2011be,Chu:2013jja,Blennow:2013jba,Bernal:2017kxu,DEramo:2017ecx}, DM is gradually produced via “energy leakage” from the thermal bath of SM particles. The couplings required to reproduce the observed DM abundance are then many orders of magnitude smaller than for the freeze-out mechanism, complying with existing data and offering promising targets for future experiments~\cite{Co:2015pka,Hambye:2018dpi,Belanger:2018sti,Heeba:2019jho,No:2019gvl,Caloni:2024olo,Ghosh:2024nkj,Yin:2024sle}. Over recent years, calculations of freeze-in production have reached a high level of sophistication, with automated tools~\cite{Belanger:2018ccd,Bringmann:2021sth} including thermal corrections and quantum effects~\cite{Darme:2019wpd,DeRomeri:2020wng,Biondini:2020ric,Dvorkin:2019zdi,Chakrabarty:2022bcn,Becker:2023vwd}.

The freeze-in mechanism, however, comes with two complications not present in the freeze-out picture. The first is that, since no equilibrium state is ever reached, the final abundance depends on the initial conditions, i.e.,\ the abundance of DM particles before freeze-in becomes efficient. In the literature, it is commonly assumed that this initial abundance is zero, although various works have pointed out that DM production through gravitational effects~\cite{Bernal:2020ili,Mambrini:2021zpp,Lebedev:2022ljz,Lebedev:2022cic,Zhang:2023xcd,Henrich:2024rux} or inflationary dynamics~\cite{Garcia:2020eof,Garcia:2021iag,Barman:2022tzk,Becker:2023tvd,Freese:2024ogj} may be non-negligible. The second is that in many models the freeze-in abundance is directly sensitive to the details of reheating, i.e. to the temperature when the universe first entered radiation domination~\cite{Elahi:2014fsa,Chen:2017kvz,Bernal:2019mhf}. This is in particular the case for DM production via effective operators (induced by new physics above the reheating temperature~\cite{Barman:2020plp,Frangipane:2021rtf}) and for DM particles with a mass above the reheating temperature.

These issues become particularly relevant in models with a low reheating temperature~\cite{Silva-Malpartida:2023yks,Bernal:2024ndy}, which have received considerable attention recently because they allow freeze-in with comparably large couplings~\cite{Cosme:2023xpa,Boddy:2024vgt,Arcadi:2024wwg,Cox:2024oeb,Mondal:2025awq}. A common assumption in these models is that inflationary reheating is followed by a second period of reheating, in which the universe deviates from radiation domination and the entropy of the SM thermal bath is not conserved. The simplest realization of this idea is to consider an out-of-equilibrium matter component that slowly decays into relativistic SM particles after an early period of matter domination~\cite{Co:2015pka,Calibbi:2021fld,Banerjee:2022fiw,Silva-Malpartida:2024emu}. Alternative possibilities for the equation of state of the universe and the temperature dependence of the energy transfer have been explored in several recent studies~\cite{Cosme:2024ndc,Bernal:2025fdr}. However, these studies typically assume that the final DM abundance is determined exclusively by the second period of reheating and that any earlier contributions can be neglected (see however ref.~\cite{Allahverdi:2019jsc} for a notable exception).

In the present work, we study this assumption in a general way by considering a scalar field that undergoes a first-order phase transition in the early universe.\footnote{The impact of the electroweak phase transition on the freeze-in production of DM was previously explored in refs.~\cite{Heeba:2018wtf,Lebedev:2019ton}. Other studies have studied the impact of a dark sector phase transition on how DM particles can be produced or destroyed~\cite{Baker:2017zwx,Wong:2023qon,Ayazi:2025sfs}.} Following inflationary reheating, the scalar field sits at a false minimum with sizeable vacuum energy, from which it transitions to the true minimum at some lower temperature. After the phase transition, the oscillations of the scalar field around the true minimum decay into SM particles, leading to a second period of reheating. This set-up comprises the case of early matter domination, but also allows for the possibility that the scalar field dominates the energy density of the universe before the phase transition, leading to a period of accelerated expansion. In this case, it is possible that the SM temperature increases rapidly after the phase transition, provided the scalar field oscillations decay sufficiently quickly.

We derive analytical approximations that describe the various stages of cosmological evolution, which we validate against numerical solutions of the coupled system of Boltzmann equations. We then determine the conditions that must be fulfilled in order for the final abundance of DM to be determined primarily by the details of the phase transition and the subsequent reheating and to be insensitive to the details of inflationary reheating. We refer to this scenario as \emph{DM phase-in}, both because DM production happens in several phases and because the dominant contribution follows from a first-order phase transition.

Another intriguing possibility is that the freeze-in contribution (from inflationary reheating) and the phase-in contribution (from the second period of reheating) are of comparable magnitude. In this case, our set-up predicts that the two populations would have substantially different temperatures, such that a mixture of warm and cold DM can be achieved from a single particle species. We discuss possible implications for Lyman-alpha forest data and the 21-cm signal~\cite{DEramo:2020gpr,Du:2021jcj,Decant:2021mhj,Xu:2024uas}.

The remainder of this work is structured as follows. In section~\ref{sec:set-up}, we discuss the general set-up that we consider and define the different stages of cosmological evolution. In section~\ref{sec:evolution}, we provide the relevant Boltzmann equations and present approximate analytical solutions, which we compare with numerical results in section~\ref{sec:comparison}. In section~\ref{sec:phase-in}, we finally determine the regions of parameter space that correspond to DM phase-in, followed by a discussion of possible implications for observations in section~\ref{sec:pheno}. Our conclusions are presented in section~\ref{sec:conclusions}. Appendix~\ref{app:RD} provides additional details on our calculations. 

\section{General set-up}
\label{sec:set-up}

In this study, we consider a universe that -- after inflation and the subsequent reheating -- is filled with four different energy components: a bath of relativistic SM particles, a bath of relativistic dark sector particles, i.e.\ dark radiation (DR), a scalar field $\phi$ and a DM species.  We denote the corresponding energy densities by $\rho_\text{SM}$, $\rho_\text{DR}$, $\rho_\phi$ and $\rho_\text{DM}$, respectively. We do not consider the details of primordial reheating and simply start our discussion at the reheating temperature $T_\mathrm{RH}$, i.e.\ the highest temperature at which the universe is dominated by radiation.\footnote{We emphasize that the maximal temperature $T_\mathrm{max}$ of the universe can be higher than the temperature  $T_\mathrm{RH}$ at which the universe first enters radiation domination~\cite{Giudice:2000ex}. However, it has been shown that in most cases of interest, the freeze-in yield depends only on $T_\text{RH}$ and not on $T_\text{max}$~\cite{Bernal_2019, Cox:2024oeb}. We will return to this issue in section~\ref{subsec:evolution_III}.}

We assume that the reheating temperature is large enough that all SM particles are in equilibrium with each other and can be described by a common temperature $T$, such that
\begin{equation}
\label{eq:rho_SM}
\rho_\mathrm{SM} = \frac{\pi^2}{30} g_{\star} T^4
\end{equation}
with $g_\star$ denoting the effective number of relativistic degrees of freedom.
Likewise, the energy density of dark radiation is given by
\begin{equation}
\label{eq:rho_DR}
\rho_\mathrm{DR} = \frac{\pi^2}{30} g_\text{DR} T_\text{DR}^4 \; .
\end{equation}
We assume that the two radiation baths are extremely weakly coupled, such that their temperatures can in principle be different: $T_\text{DR} \neq T$. The precise value of $T_\text{DR}$ plays however no role in the subsequent discussion, so that we set $T_\text{DR} = T$ for simplicity. We furthermore assume that $g_\text{DR}$ is small enough that the energy density of dark radiation only gives a tiny contribution to the total energy density, i.e.\ that $\rho_\mathrm{DR} \ll \rho_\mathrm{SM}$. For concreteness, we set $g_\text{DR} = 2$, as appropriate for example for a gauge field.

Nevertheless, the dark radiation plays an important role because it strongly couples to the scalar field $\phi$ and gives rise to thermal corrections that create a temperature-dependent effective potential $V(\phi, T)$ and determine the evolution of the scalar field. Since we are interested in describing the impact of a first-order phase transition on freeze-in in a general manner, we do not specify the details of these interactions and the resulting potential, but simply assume that the scalar field initially sits in a metastable vacuum of the scalar potential, such that its energy density is constant and given by the latent heat 
\begin{equation}
\label{eq:rho_phi}
    \rho_\phi=\Delta V \; ,
\end{equation}
where $\Delta V$ is the potential energy difference between the false and true minima. As the universe cools down, the potential barrier separating the two minima decreases, and the scalar field can transition to the true minimum through the nucleation, expansion and collision of bubbles.

The DM component is assumed to be very feebly coupled to both the SM and the DR thermal baths and never enters into equilibrium with either of them. We also assume that primordial reheating does not produce DM, such that $\rho_\text{DM}(T_\text{RH}) = 0$. Subsequently, DM particles are produced from the SM thermal bath via the freeze-in mechanism. We do not specify the details of the production process, but simply assume that it proceeds via a non-renormalisable operator of dimension $4+n$ with $n>0$. If the mass of the DM particles is negligible compared to $T$, the production cross section can on dimensional grounds be written as
\begin{equation}
    \sigmaV = \frac{T^{2(n-1)}}{\Lambda^{2n}}\; ,
    \label{eq:DM_production}
\end{equation}
where $\Lambda$ is the energy scale at which the interaction is generated, see also ref.~\cite{Elahi:2014fsa}.

The Hubble rate is given by the Friedmann's equation,
\begin{equation}
    H=\frac{\dot a}{a}=\sqrt{\frac{8 \pi}{3 M_\mathrm{Pl}^2}(\rho_\mathrm{SM} + \rho_{\phi} + \rho_\mathrm{DR})}\; ,
    \label{eq: hubble rate}   
\end{equation}
where $ M_\mathrm{Pl} = 1.22 \times 10^{19}\,$GeV is the Planck mass, $a$ is the scale factor and $\rho_\mathrm{tot} = \rho_\mathrm{SM} + \rho_{\phi} + \rho_\mathrm{DR}$. The DM contribution $\rho_\mathrm{DM}$ to the dynamics of the expansion can be ignored, since its energy density is very small in the early universe. 

The subsequent evolution of the universe can be divided into several stages:
\begin{itemize}
 \item \textbf{Stage I:} The universe is radiation dominated, such that $H \propto T^2$ and the entropy of the SM thermal bath is conserved (neglecting the tiny amount of DM production), such that $T \propto a^{-1}$.
 \item \textbf{Stage II:} As the universe cools down, the vacuum energy of the scalar field will start to dominate the energy density of the universe, leading to a period of accelerated expansion, which will rapidly deplete and cool down the SM and DR thermal baths. This in turn triggers the first-order phase transition of the scalar field to the true minimum.
\end{itemize}

The redshift corresponding to the transition from stage I to stage II, i.e. the beginning of vacuum domination (VD), can be calculated from the condition $\rho_\phi(a_\mathrm{VD})\approx\rho_\mathrm{SM} (a_\mathrm{VD})$ as
\begin{equation}
          a_\mathrm{VD} \approx  a_\mathrm{RH} T_\mathrm{RH}\left(\frac{\pi^2}{30} \frac{g_{\star}}{\Delta V}\right)^{1/4} \; .
\end{equation}
The temperature $T_\text{PT}$ of the phase transition can in principle be calculated from $V(\phi, T)$, but in our simplified approach we simply take it as a free parameter. The corresponding redshift is given by
    \begin{equation}
    \label{eq:aPT}
        a_\mathrm{PT} \approx a_\mathrm{RH}\left(\frac{T_\mathrm{RH}}{T_\mathrm{PT}}\right) \; .
    \end{equation}

During the phase transition, the vacuum energy present in stage II is quickly converted into oscillations of the scalar field around the true minimum. Moreover, the vacuum expectation value of the scalar field gives mass to the DR particles, which rapidly decay or annihilate into scalar field excitations (see refs.~\cite{Bringmann:2023iuz,Balan:2025uke} for concrete examples of this set-up). While the scalar field still dominates the energy density of the universe, it no longer behaves like vacuum energy, but instead has an energy density that decreases as the universe expands. Moreover, the scalar field excitations can decay into SM particles with a characteristic timescale $\Gamma$, which we assume to be constant for simplicity.\footnote{We emphasize that it is plausible that decays of the scalar field only become relevant after the phase transition, for example if the phase transition spontaneously breaks a discreet or continous symmetry that stabilises the scalar field before the phase transition.} This leads to the following stages of evolution:
\begin{itemize}
 \item \textbf{Stage III:} The universe is dominated by the scalar field $\phi$, which slowly decays into SM particles. As a result, the entropy of the SM thermal bath is no longer conserved and the temperature decreases more slowly than $a^{-1}$. Eventually, the energy density of the SM thermal bath becomes comparable to the energy density of the scalar field, which then quickly becomes negligible.
 \item \textbf{Stage IV:} The universe once again enters into radiation domination and remains there for the subsequent stages of cosmological evolution, such as Big Bang Nucleosynthesis.
\end{itemize}

In the literature, it is often assumed that the scalar field oscillations after a first-order phase transition can be treated as coherent, i.e., they behave approximately like matter with pressure $p_\phi = 0$, which would lead to a period of early matter domination if the decay of the scalar field is slow (see, for example, \cite{Ellis:2019oqb, Ellis:2020nnr}). However, given the highly inhomogeneous state of the universe during bubble nucleation and collision, it is far from clear that this assumption holds. We therefore allow for a general equation of state for the scalar field: 
\begin{equation}
p_\phi = \omega \rho_\phi 
\label{eq:eos}
\end{equation}
with $\omega = 0$ ($\omega = 1/3$) corresponding to matter (radiation). The energy density of the scalar field then scales as $\rho_\phi \propto a^{-3(1+\omega)}$ (as long as decays can be neglected) and the Hubble rate as $H \propto a^{-3(1+\omega)/2}$.

As we will see below, stage III ends approximately when the Hubble rate is equal to the decay rate $\Gamma$. This relation can be used to estimate the value of the scale factor when the universe transitions to radiation domination:
\begin{equation}
    \label{eq:aRD_approx}
 a_\text{RD}^\text{approx} = a_\text{PT} \left( \frac{\sqrt{8 \pi \Delta V / 3}}{\Gamma M_\text{Pl}} \right)^{\tfrac{2}{3(1+\omega)}} \; .
\end{equation}
The temperature of the SM thermal bath at this point can be obtained from the relation $\rho_\phi(a_\text{RD}) \approx \rho_\text{SM}(a_\text{RD})$ as
\begin{equation}
\label{eq:TRD_approx}
T_\text{RD}^\text{approx} = \left(\frac{45}{4 g_\star \pi^3 }\right)^{1/4} \sqrt{\Gamma M_\text{Pl}} \; .
\end{equation}
As shown in appendix~\ref{app:RD}, these estimates can be refined further in order to obtain a better approximation of the full numerical results. The resulting expressions are
\begin{align}
    \label{eq:aRD}
        a_\mathrm{RD} & = a_\mathrm{PT} \left(\sqrt{\frac{2 \pi}{3}}\frac{(5-3\omega) \sqrt{\Delta V}}{\Gamma M_\mathrm{Pl}}\right)^{\frac{2}{3(1+\omega)}} \; , \\
    \label{eq:TRD}
        T_\mathrm{RD} & =\left(\frac{45 }{g_\star\pi^3}\right)^{1/4}\sqrt{\frac{\Gamma M_\mathrm{Pl}}{(5-3\omega)}} \;.
\end{align}
We will use these more accurate expressions in the following.

\begin{figure}
    \centering
    \includegraphics[width=0.95\linewidth]{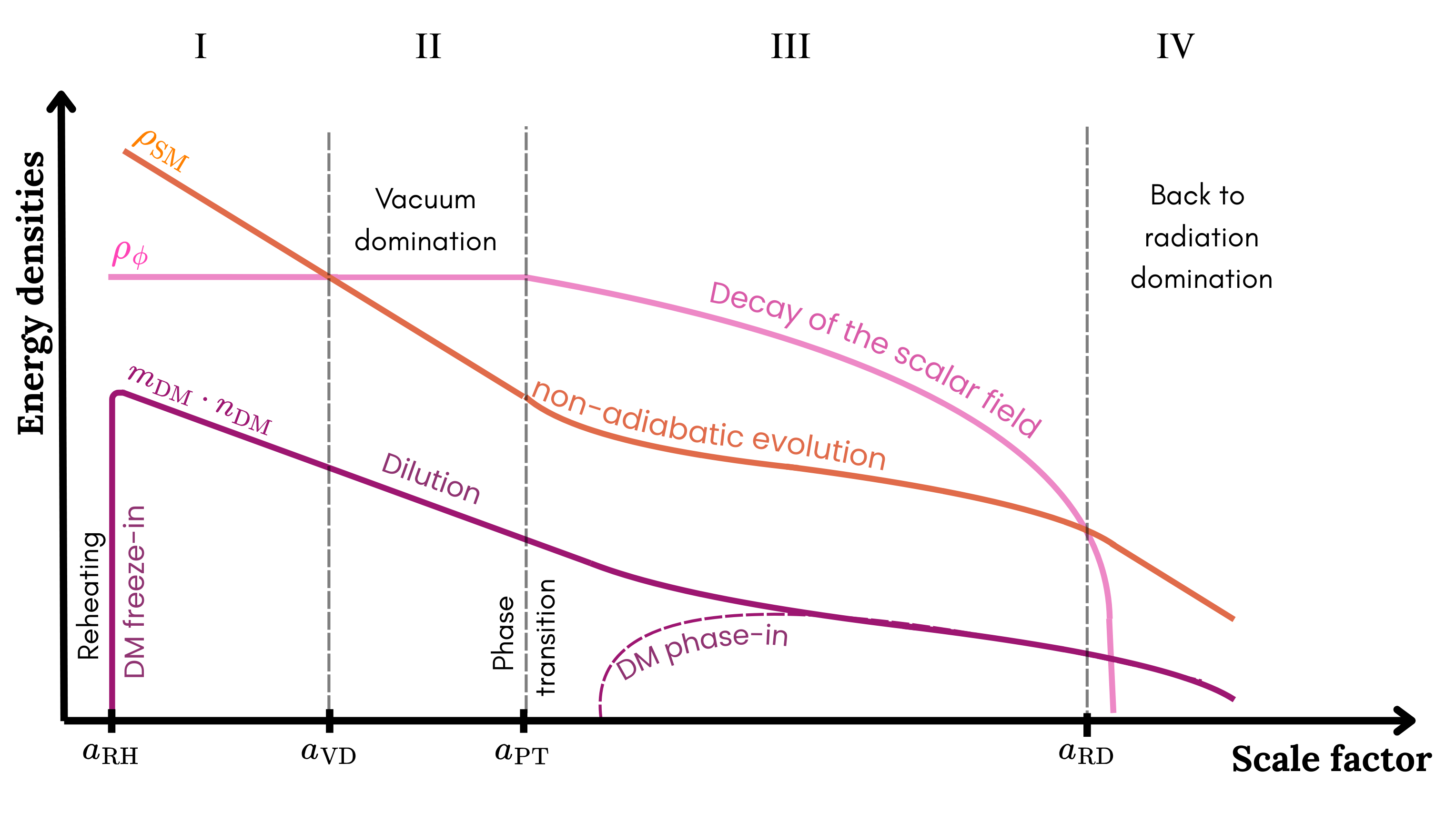}
    \caption{Sketch of the evolution of the energy densities of the scalar field (pink), SM radiation bath (orange) and dark matter (purple) during the four stages introduced in section~\ref{sec:set-up}. For simplicity, the dark radiation energy density is omitted. At the end of inflationary reheating (RH), the universe is initially radiation dominated and DM is produced via UV freeze-in. A supercooled phase transition (PT) takes place after a period of vacuum domination (VD). After the phase transition, the scalar field behaves as a fluid with an equation of state parameter $\omega \geq 0$. The subsequent decay of the scalar field leads to a second reheating of the SM particles and to an extra contribution to the DM number density. This late production is referred to as \textit{phase-in}. Finally, the universe is radiation dominated (RD) again.}
    \label{fig:sketch}
\end{figure}

The evolution of the energy densities as a function of scale factor is illustrated in figure~\ref{fig:sketch}.
To summarize, our set-up is described by the following parameters:
\begin{itemize}
 \item $T_\text{RH}$ and $\Delta V$ describing the initial conditions, see eqs.~\eqref{eq:rho_SM} and~\eqref{eq:rho_phi};
 \item $T_\text{PT}$ and $\Gamma$ characterising the transition from stage II to stage III (see eq.~\eqref{eq:aPT}) and from stage III to stage IV (see eq.~\eqref{eq:aRD}), respectively;
 \item $n$ describing the temperature dependence of DM production, see eq.~\eqref{eq:DM_production};
 \item $\omega$ determining the equation of state of the scalar field oscillations, see eq.~\eqref{eq:eos}.
\end{itemize}
In order to calculate the DM abundance $\Omega_\text{DM} h^2$, we also need to specify the new physics scale $\Lambda$ that enters in the DM production rate and the DM mass $m_\text{DM}$ that is needed to convert from the number density to the energy density of a non-relativistic species. However, as long as the DM mass is small compared to the relevant temperature scales, the relation is quite simple: $\Omega_\text{DM} h^2 \propto m_\text{DM} / \Lambda^{2 n}$. In particular, these parameters drop out when calculating \emph{relative} changes to the DM relic abundance.

\section{Detailed evolution: Analytical approximations}
\label{sec:evolution}

Having introduced the general features of the model that we consider, we now discuss in detail the relevant Boltzmann equations for each stage of the evolution and provide approximate analytical solutions. 

\subsection{Stage I: Post-inflationary radiation domination}

In the first two stages of the evolution, the differential equations describing the energy densities of the SM thermal bath and the scalar field are simply given by
    \begin{align}
            \frac{\dd \rho_\phi}{\dd a}&= 0 \; ,\\
            \frac{\dd \rho_\mathrm{SM}}{\dd a}&= -\frac{4}{a} \rho_\mathrm{SM} \; .
    \end{align}
In other words, the scalar energy density is constant, while the SM radiation energy densities decreases proportional to $a^{-4}$. An analogous equation holds for the DR energy density but is not relevant for the further discussion.

To study the production of DM via the freeze-in mechanism, it is convenient to consider the Boltzmann equation for the DM number density:
\begin{equation}
    \frac{\mathrm{d} n_\mathrm{DM}}{\mathrm{d} a}= -\frac{3}{a} n_\mathrm{DM} +\frac{\sigmaV}{a H} n_\mathrm{SM}^2 \label{eq:nDM_evolution} \; ,
\end{equation}
where $n_\mathrm{SM}(T) = \frac{\zeta(3)}{\pi^2} g_n T^3$ is the number density of SM particles with $g_n$ denoting the relevant degrees of freedom.\footnote{In a realistic model, only some of the SM particles will participate in the production of DM particles. As long as all relevant particles are relativistic, this changes the production rate only by a constant factor, which can be absorbed into an appropriate redefinition of $\Lambda$.}

To obtain a simple analytical estimate for $n_\mathrm{DM}$, we can approximate the Hubble rate as being determined solely by the energy density of the SM thermal bath
\begin{equation}
    H \approx \sqrt{\frac{4 \pi^3 g_\ast}{45}} \frac{T_\mathrm{RH}^2}{M_{Pl}} \left( \frac{a_\mathrm{RH}}{a}\right)^2 \; .
\end{equation}
In this case, the Boltzmann equation can be rewritten as
\begin{equation}
    \frac{\dd n_\mathrm{DM}}{\dd a}= -\frac{3}{a} n_\mathrm{DM} +\frac{\left(n_\mathrm{SM}^\text{RH}\right)^2}{\Lambda^{2n}} \sqrt{\frac{45}{4 \pi^3  g_\star}}  
    \frac{a_\mathrm{RH}^{2(n+1)}}{a^{2n+3}} M_\mathrm{Pl} T_\mathrm{RH}^{2(n-2)} \; ,
\end{equation}
where $n_\mathrm{SM}^\text{RH}$ is the SM number density at reheating. Eq.~\eqref{eq:nDM_evolution} then yields
\begin{equation}
    n_\mathrm{DM}^\text{I}(a)= \frac{n_\mathrm{SM}^2(T_\text{RH})}{\Lambda^{2n}} \sqrt{\frac{45 M_\mathrm{Pl}^2}{4 \pi^3  g_\star}} \, \frac{T_\mathrm{RH}^{2(n-2)}}{2n-1} \times  \left[ \left(\frac{a_\mathrm{RH}}{a}\right)^3 - \left(\frac{a_\mathrm{RH}}{a}\right)^{2(n+1)} \right] \; ,
\end{equation}
which for $n \geq 1$ and $a \gg a_\text{RH}$ can be approximated as
\begin{equation}
    n_\mathrm{DM}^\text{I}(T) = \left( \frac{\zeta(3) g_n}{\pi^2} \right)^2 \sqrt{\frac{45}{4 \pi^3  g_\star}} \, \frac{M_\text{Pl}}{\Lambda^{2n}} \, \frac{T_\mathrm{RH}^{2n-1}}{2n-1} \, T^3 \propto \frac{T_\mathrm{RH}^{2n-1} \, T^3 \, M_\text{Pl}}{\Lambda^{2n}} \ \; .
\end{equation}
This is the well-known result for ultraviolet freeze-in: the comoving DM number density $n_\text{DM} a^3$ is sensitive to $T_\text{RH}$ and becomes independent of $T$ for $T \ll T_\text{RH}$. The total amount of DM produced in stage I is obtained by evaluating $n_\text{DM}$ at $T = T_\text{VD}$. Since in our set-up, DM particles are only produced and never destroyed, this abundance will remain until the present day. Its contribution to the total DM density at later times is simply given by a dilution factor proportional to $(a_\text{VD} / a)^3$.

\subsection{Stage II: Vacuum domination and phase transition}

The differential equations describing this stage are the same as for the previous one. However, the Hubble rate is now dominated by the energy density of the scalar field, such that
\begin{equation}
    H \approx \sqrt{\frac{8 \pi \Delta V }{3 M_{Pl}^2}}
\end{equation}
which leads to a different scaling of the DM number density:
\begin{equation}
    n_\mathrm{DM}^\text{II}(a) = \frac{n_\mathrm{SM,VD}^2}{\Lambda^{2n} H}  \, \frac{T_\mathrm{VD} ^{2(n-1)}}{2n+1} \times  \left[ \left(\frac{a_\mathrm{VD}}{a}\right)^3 - \left(\frac{a_\mathrm{VD}}{a}\right)^{4 + 2n} \right] \; .
\end{equation}
For $a \gg a_\text{VD}$ this leads to
\begin{equation}
    n_\mathrm{DM}^\text{II}(T) \propto \frac{T_\mathrm{VD}^{2n+1} \, M_\text{Pl} \, T^3}{\Lambda^{2n} \sqrt{\Delta V}} \propto \frac{T_\mathrm{VD}^{2n-1} \,  M_\text{Pl} \, T^3}{\Lambda^{2n}} \; ,
\end{equation}
where in the last step we have used that $\rho_\text{SM}(T_\text{VD}) = \Delta V$. It follows that for $T_\text{VD} \ll T_\text{RH}$ the amount of DM produced during the second stage is negligible compared to the production during stage I.

\subsection{Stage III: Reheating through scalar field decays}
\label{subsec:evolution_III}
After the phase transition, the evolution of the scalar field and the SM thermal bath are given by the Boltzmann equations
\begin{align}
    \frac{\mathrm{d} \rho_\phi}{\mathrm{d} a} & = -\frac{3 (1+\omega)}{a} \rho_{\phi} - \frac{\Gamma}{a H} \rho_{\phi} \, ,\label{eq:phi_evolution_APT} \\
    \frac{\mathrm{d} \rho_\mathrm{SM}}{\mathrm d a} & = -\frac{4}{a} \rho_\mathrm{SM} +\frac{\Gamma}{a H} \rho_\phi \; ,\label{eq:SM_evolution_APT}
\end{align}
which implicitly define the decay rate $\Gamma$.
The energy injection term proportional to $\Gamma$ implies that the SM radiation energy density no longer evolves adiabatically, i.e.\ it no longer decreases as $\rho_\mathrm{SM}\propto a^{-4}$. 
    
To obtain an approximate analytical solution, we make use of the fact that $\Gamma < H$ during stage III, such that we can neglect the second term on the right-hand side of eq.~\eqref{eq:phi_evolution_APT}:
\begin{align}
            \frac{\dd \rho_\phi}{\dd a}& \approx -\frac{3 (1+\omega)}{a} \rho_{\phi} \; .
            \label{eq:phi_stageIII}
\end{align}
Using furthermore that the energy density of the scalar field gives the dominant contribution to the Hubble rate, we obtain
\begin{equation}
    H \approx \sqrt{\frac{8 \pi \Delta V }{3 M_\text{Pl}^2}} \left(\frac{a_\mathrm{PT}}{a}\right)^{3(1+\omega)/2} \; ,
\end{equation}
which we can use to solve eq.~\eqref{eq:SM_evolution_APT} and obtain \begin{align}
    \rho_\mathrm{SM}^\text{III}(a) & = \frac{\pi^2}{30}g_{*} \left(\frac{a_\mathrm{PT}}{a}\right)^4\left( T_\mathrm{PT}^4  + \frac{\Gamma M_\mathrm{Pl}}{(5- 3 \omega)}\frac{60}{\pi^2g_*}\sqrt{\frac{3\Delta V}{8 \pi}}\left[\left(\frac{a}{a_\mathrm{PT}}\right)^{5/2-3\omega/2} -1 \right] \right) \nonumber \\
    & = \frac{\pi^2}{30}g_{*} \left(\frac{a_\mathrm{PT}}{a}\right)^4 \left(T_\mathrm{PT}^4   + T_\Gamma^4 \left[\left(\frac{a}{a_\mathrm{PT}}\right)^{5/2-3\omega/2} -1 \right] \right) \label{eq:rhoSM_APT}
\end{align}
where in the second line we have defined
\begin{equation}
T_\Gamma^4 \equiv \frac{\Gamma M_\mathrm{Pl}}{(5- 3 \omega)}\frac{60}{\pi^2g_\star}\sqrt{\frac{3\Delta V}{8 \pi}} \; 
\label{eq:Teff}
\end{equation}
to simplify notation.
The first term in eq.~\eqref{eq:rhoSM_APT} corresponds to the pre-existing SM energy density, while the second term corresponds to the one produced through scalar field decays.

For a supercooled phase transition with $\Delta V \gg \rho_\text{SM}^\text{PT}$ and a sufficiently slow scalar field decay (such that $a_\text{RD} \gg a_\text{PT}$) the second term dominates as $a \to a_\text{RD}$, such that we can approximate 
\begin{equation}
    n_\mathrm{SM}(a) = \frac{\zeta(3)}{\pi^2}g_n T_\Gamma^3 \left(\frac{a}{a_\mathrm{PT}}\right)^{\tfrac{-9 (1+\omega)}{8}} \; .
\end{equation}
Using this expression for $n_\text{SM}(a)$, the additional contribution to the DM freeze-in from stage III is found to be
\begin{equation}
     n_\mathrm{DM}^\text{III}(a) = T_\Gamma^{4+2n}  \left(\frac{\zeta(3) g_n}{\Lambda^n\pi^2} \right)^2 \sqrt{\frac{2}{3 \pi \Delta V}}\frac{M_\text{Pl}}{4-n( \omega+1)}\left[\left(\frac{a_\mathrm{PT}}{a}\right)^{(\omega+1)3n/4} - \left(\frac{a_\mathrm{PT}}{a}\right)^3 \right]
\end{equation}
for $n \neq 4 / (\omega + 1)$. In the following, we will focus on $n < 4 / (\omega + 1)$, such that the first term in the final bracket dominates for $a \gg a_\text{PT}$. Re-substituting $T_\Gamma$ from eq.~\eqref{eq:Teff} and evaluating the resulting expression at $a_\text{RD}$ as given in eq.~\eqref{eq:aRD} yields
\begin{align}
     n_\mathrm{DM}^\text{III}(a_\text{RD}) & = \frac{30 \zeta(3)^2 g_n^2}{\pi^7 g_\star[4 - n (1 + \omega)]} \left(\frac{45}{\pi^3 g_\star}\right)^{n/2} \frac{M_\text{Pl} \, (M_\text{Pl} \Gamma)^{1 + n}}{ (5 - 3 w)^{(1+n)}\Lambda^{2 n}} \quad \propto \frac{T_\text{RD}^{2(1+n)} M_\text{Pl}}{\Lambda^{2n}}
\end{align}
with $T_\text{RD}$ given in eq.~\eqref{eq:TRD}. Interestingly, we find that the final result is independent of the temperature of the phase transition and the initial energy density of the scalar field. This finding generalises the known result that for an early period of matter domination ($\omega = 0$) freeze-in is UV-insensitive for $n < 4$~\cite{Co:2015pka,Chen:2017kvz,Garcia:2017tuj}.

\subsection{Stage IV: Return to radiation domination}

DM production during stage IV can be calculated in complete analogy to stage I. Since the universe is now once again in a period of radiation domination, freeze-in production is UV-dominated, i.e. sensitive to $T_\text{RD}$. We find for $a \gg a_\text{RD}$:
\begin{equation}
    n_\mathrm{DM}^\text{IV}(a)\approx \frac{n_\mathrm{SM,RD}^2}{\Lambda^{2n}} \sqrt{\frac{45 M_\mathrm{Pl}^2}{4 \pi^3  g_{*}}} \, \frac{T_\mathrm{RD}^{2(n-2)}}{2n-1} \left(\frac{a_\mathrm{RD}}{a}\right)^3 \propto \frac{T_\mathrm{RD}^{2n-1} \, T^3 \, M_\text{Pl}}{\Lambda^{2n}} \ \; 
    \label{eq:nDM_IV}.
\end{equation}
We will see that the late contributions from stages III and IV can constitute a significant part of the final DM abundance even if $a_\mathrm{RD}$ is orders of magnitude larger than $a_\mathrm{RH}$.

\subsection{Combining all contributions}

From the previous results, the total DM density at late times i.e $a \gg a_\text{RD}$ is given by
\begin{equation}
    n_\text{DM}^\text{tot}(a) = n_\text{DM}^\text{I}(a_\text{VD}) \left(\frac{a_\text{VD}}{a}\right)^3 + n_\text{DM}^\text{II}(a_\text{PT}) \left(\frac{a_\text{PT}}{a}\right)^3 + n_\text{DM}^\text{III}(a_\text{RD}) \left(\frac{a_\text{RD}}{a}\right)^3 + n_\text{DM}^\text{IV}(a) \; .
\end{equation}
Since the entropy of the SM thermal bath is conserved in stages II and IV, it follows that $a_\text{VD} / a_\text{PT} = T_\text{PT} / T_\text{VD}$ and $a_\text{RD} / a = T / T_\text{RD}$. However, during stage III the total entropy $S$ of the SM changes by a factor
\begin{equation}
\label{eq:dilution}
 D = \frac{S_\text{RD}}{S_\text{PT}} 
= \left(\frac{T_\text{RD} a_\text{RD}}{T_\text{PT} a_\text{PT}}\right)^3 = 
\left(\frac{90}{8 \pi^3 g_\star}\right)^{3/4} \left(\frac{2\, \Gamma M_\text{Pl}}{5-3 \omega}\right)^{\tfrac{3}{2} - \tfrac{2}{1+\omega}} \left(\frac{8 \pi \Delta V}{3} \right)^{\tfrac{1}{1+\omega}} T_\text{PT}^{-3} \; .
\end{equation}
Using this dilution factor, we can rewrite the total number density as a function of temperature as
\begin{equation}
 n_\text{DM}^\text{tot}(T) = \frac{1}{D} \left[n_\text{DM}^\text{I}(a_\text{VD}) \left(\frac{T}{T_\text{VD}}\right)^3 + n_\text{DM}^\text{II}(a_\text{PT}) \left(\frac{T}{T_\text{PT}}\right)^3\right] + n_\text{DM}^\text{III}(a_\text{RD}) \left(\frac{T}{T_\text{RD}}\right)^3 + n_\text{DM}^\text{IV}(T) \; .
 \end{equation}
As discussed above, the contribution from stage II can be neglected for $T_\text{VD} \ll T_\text{RH}$. 
The contributions from stage III and stage IV are both found to be proportional to $T_\text{RD}^{2n-1}$, i.e.\ we can write
\begin{equation}
    n_\text{DM}^\text{III}(a_\text{RD}) \left(\frac{T}{T_\text{RD}}\right)^3 = \kappa n_\text{DM}^\text{IV}(T)
\end{equation}
with
\begin{equation}
   \kappa = \frac{4 (2 n - 1)}{3(4 - n(1 + \omega))} \left(\frac{5 - 3 w}{2}\right)^{n/2}
\end{equation}
depending only on $n$ and $\omega$. For $n = 1$ and $0 \leq \omega \leq 1/3$, we find $\kappa$ around 0.7, while for $n > 1$ it becomes close to unity.

The total DM abundance can therefore be written as
\begin{equation}
 n_\text{DM}^\text{tot}(T) \approx \left(\frac{\zeta(3) g_\star}{\pi^2}\right)^2 \sqrt{\frac{45}{4 \pi^3 g_n}} \frac{T^3 M_\text{Pl}}{\Lambda^{2n}} \left[ \frac{T_\text{RH}^{2n-1}}{D} + (1 + \kappa) {T_\text{RD}^{2n-1}} \right] \; .
 \label{eq:DMtot}
\end{equation}
In the following, we will refer to the first term in the square bracket as the \emph{freeze-in contribution} to the DM density, and to the second term as the \emph{phase-in contribution}.
From this result, the present-day abundance of DM can be calculated using entropy conservation:
\begin{equation}
    \Omega_\text{DM} h^2 = \frac{m_\text{DM}  \, n_\text{DM}^\text{tot}(T_\text{end})}{s(T_\text{end})} \frac{s_0}{\rho_{c,0} / h^2} \; ,
    \label{eq: DM relic}
\end{equation}
where $h = H_0 / (100 \, \mathrm{km \, s^{-1} \, Mpc^{-1}} \approx 0.68$, $s(T)$ denotes the SM entropy density, $s_0$ its present-day value and $\rho_{c,0}$ the critical density of the present universe. The temperature  $T_\text{end}$ should be sufficiently smaller than $T_\text{RD}$ but is otherwise arbitrary.

\subsection{Instantaneous scalar decays}

In the discussion above we have assumed that the scalar field decays slowly, i.e.\ that immediately after the phase transition $\Gamma \ll H$. However, it is also conceivable that $\Gamma$ is so large that the scalar field oscillations decay immediately after the phase transition. In this case, all of the energy stored in the scalar field is rapidly transferred to the SM thermal bath, such that $a_\text{PT} = a_\text{RD}$ and stage III is absent. For a strongly supercooled phase transition, the temperature at the beginning of stage IV is then given by $\rho_\text{SM}(T_\text{RD}) = \Delta V$ and hence
\begin{equation}
  T_\text{RD}^\text{inst} =  \left( \frac{30 \Delta V}{\pi^2 g_\star} \right)^{1/4}
\end{equation}
independent of $\Gamma$ and $\omega$. Using this expression for $T_\text{RD}$, the DM density produced during stage IV is again given by eq.~\eqref{eq:nDM_IV}. 

We can write the total DM abundance produced in this case in the same form as eq.~\eqref{eq:DMtot} with $\kappa$ replaced by $\kappa^\text{inst} = 0$ and $D$ replaced by
\begin{equation}
    D^\text{inst} = \left(\frac{T_\text{RD}^\text{inst}}{T_\text{PT}}\right)^3 \; .
\end{equation}

\section{Detailed evolution: Comparison with numerical results}
\label{sec:comparison}

 At first sight, some of the approximations made in the previous section may seem rather crude. However, in this section, we demonstrate that the resulting analytical expressions provide an accurate estimate of both the overall evolution of the system and the final DM abundance calculated numerically.
 
 The numerical results correspond to the full solution of the following system of equations:   
 \begin{align}
    \frac{\mathrm{d} \rho_\phi}{\mathrm{d} a} & = -\frac{3 (1+\omega)}{a} \rho_{\phi} - \frac{\Gamma}{a H} \rho_{\phi} \; ,  \label{eq:Boltzmann_Eq_phi}\\
    \frac{\mathrm{d} \rho_\mathrm{SM}}{\mathrm d a} & = -\frac{4}{a} \rho_\mathrm{SM} +\frac{\Gamma}{a H} \rho_\phi \; , \label{eq:Boltzmann_Eq_SM}\\
    \frac{\mathrm{d} n_\mathrm{DM}}{\mathrm{d} a} &= -\frac{3}{a} n_\mathrm{DM} +\frac{\sigmaV}{a H} n_\mathrm{SM}^2 \;. \label{eq:Boltzmann_Eq_DM}
\end{align}
 The Hubble rate $H$ includes all contributions as defined in eq. (\ref{eq: hubble rate}). Unlike the analytical approach, the numerical solver does not differentiate between stage I and II, nor between stage III and IV. We only make the distinction between before and after the phase transition: for $a<a_\mathrm{PT}$, the scalar field behaves as a constant vacuum energy component (with $\omega=-1$ and $\Gamma=0$), while for $a>a_\mathrm{PT}$, it decays like a matter- or radiation-like fluid (with $0\leq\omega \leq 1/3$ and $\Gamma\neq 0$). To precisely determine the scale factor $a_\mathrm{PT}$ corresponding to the phase transition temperature, we track the evolution of DR until $T_\mathrm{RD}=T_\mathrm{PT}$. The transition from before to after the phase transition is defined by requiring continuity 
\begin{align}
    \rho^\text{after}_\phi(a_\text{PT}) & =     \rho^\text{before}_\phi (a_\text{PT}) + \rho^\text{before}_\text{DR} (a_\text{PT}) \;, \\
     \rho^\text{after}_\text{SM}(a_\text{PT}) & =     \rho^\text{before}_\text{SM} (a_\text{PT}) \, ,\\
    n^\text{after}_\text{DM}(a_\text{PT}) & =     n^\text{before}_\text{DM} (a_\text{PT}) \, . 
\end{align}
In the first line, we added the energy density of the dark radiation to that of the scalar field, since in our set-up we assume that these particles become heavy due to the non-zero VEV of $\phi$ and decay rapidly into scalar field excitations.

We solve the system of equations numerically up to a pre-specified scale factor $a_\text{end}$ shortly after the reestablishment of radiation domination. For $a>a_\mathrm{end}$, the SM entropy density scales as $(a_\mathrm{end}/a)^3$. However, since late-time DM production is not completely negligible, we include an analytical correction term. The DM number density today is then given by: 
\begin{equation}
\label{eq:ntoday}
    n_\mathrm{DM}^\mathrm{today}(a)= \frac{s_0}{s_\mathrm{SM}(a_\mathrm{end})} \left( n_\mathrm{DM}(a_\mathrm{end})+  n^\mathrm{late}_\mathrm{DM}\right) \; ,
\end{equation}
where  $s_0$ and $s_\mathrm{SM}(a_\mathrm{end})$ refer respectively to the SM entropy density today and after the decay of the scalar. The late-time contribution to $n_\mathrm{DM}^\text{today}$ is given by the analytical solution for freeze-in production during radiation domination (see the discussion of stage I and IV above):
\begin{equation}
     n^\mathrm{late}_\mathrm{DM}=\frac{n_\mathrm{SM,end}^2}{\Lambda^{2n}} \sqrt{\frac{45 M_\mathrm{Pl}^2}{4 \pi^3  g_{*}}} \, \frac{T_\mathrm{end}^{2(n-2)}}{2n-1} \left(1-\left(\frac{m_\mathrm{DM}}{T_\mathrm{end}}\right)^{2n-1}\right).   \; 
\end{equation}
The last term captures the kinematic suppression of DM production for $T < m_\mathrm{DM}$.

Once we include the analytical correction term for the late-time contribution to the DM density, the precise value of the endpoint $a_\text{end}$ becomes irrelevant, provided that $a_\text{end}$ is large enough to fully capture the transition to radiation domination, yet small enough to ensure that $T_\text{end} > m_\text{DM}$. In practice, we set $a_\mathrm{end} = 3 \, \text{max}( a_\mathrm{RD},a_\mathrm{PT})$ with $a_\mathrm{RD}$ given in eq.~\eqref{eq:aRD}, noting that the case $a_\mathrm{RD} < a_\mathrm{PT}$ simply means that the scalar field starts decaying immediately after the phase transition, such that $a_\mathrm{PT}$ is the relevant scale for the transition to radiation domination. We then use eqs.~\eqref{eq:ntoday} and~\eqref{eq: DM relic} to determine the relic density today. In the following, we will denote the numerical result as $\Omega^\text{num}_\text{DM} h^2$ in order to distinguish it from the analytical approximation $\Omega^\text{ana}_\text{DM} h^2$.

 \begin{figure}[t]
\centering
\begin{subfigure}{0.49\textwidth}
    \centering
    \includegraphics[width=\textwidth, scale=1]{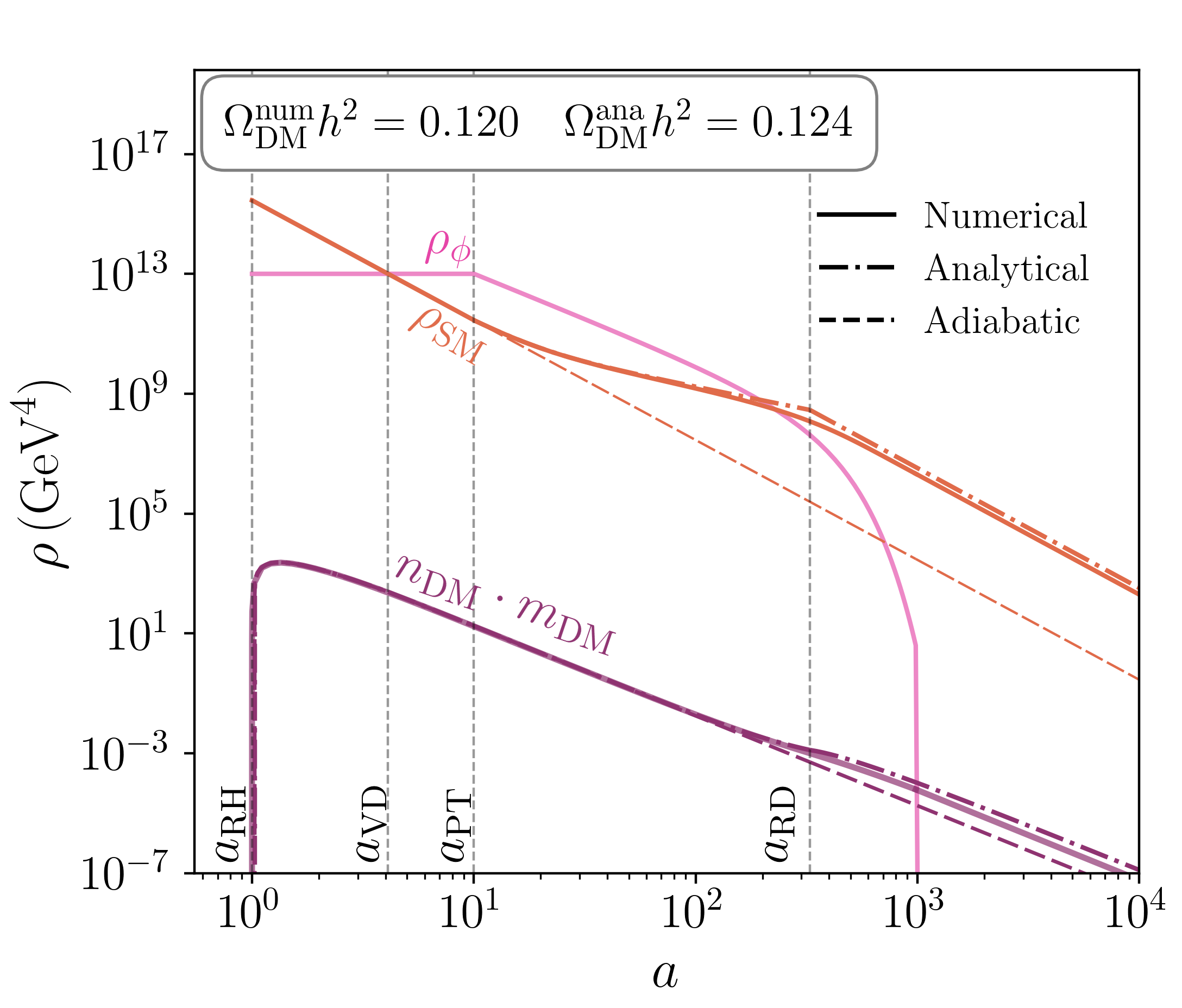}
\end{subfigure}
\begin{subfigure}{0.49\textwidth}
    \centering
    \includegraphics[width=\textwidth, scale=1]
{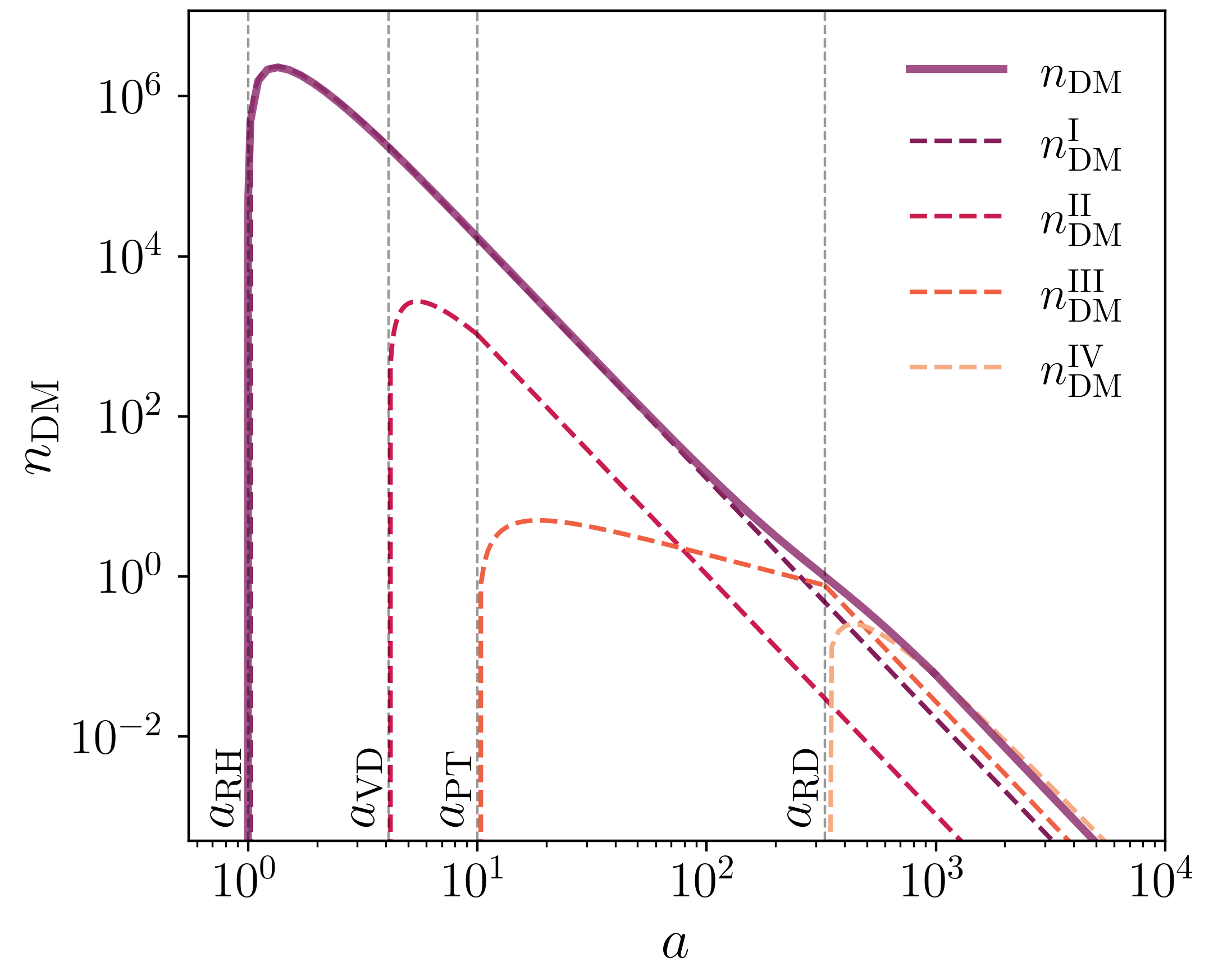}
\end{subfigure}
\caption{Comparison of the analytical solutions presented in section \ref{sec:evolution} with the full numerical results for freeze-in proceeding through a 5-dimensional operator ($n=1$) and assuming that the decaying scalar field behaves as matter ($\omega=0$). As an example benchmark point, we consider $m_\mathrm{DM}=1\,\mathrm{MeV}$, $T_\mathrm{RH}=3\cdot10^3\,\mathrm{GeV}$, $T_\mathrm{PT}=300\,\mathrm{GeV}$, $\Delta V = 10^{13}\,\mathrm{GeV}^4$, $\Gamma = 10^{-14}\,\mathrm{GeV}$ and $\Lambda=1.88 \cdot 10^{13}\,\mathrm{GeV}$. The suppression scale $\Lambda$ has been chosen such that the correct relic density of DM is produced. Left: Evolution of the energy densities of the scalar field and the SM radiation bath as well as the number density of DM as a function of the scale factor before and after the phase transition. $\Omega_\mathrm{DM}^\mathrm{num}h^2$ and $\Omega_\mathrm{DM}^\mathrm{ana}h^2$ correspond respectively to the relic density computed from the numerical and analytical results. Right: The analytical results (dashed lines) for the produced DM number density as a function of the scale factor in each of the four stages compared to the full numerical result (solid line).}
    \label{fig: numVsana_nDM}
\end{figure}

We compare the analytical and numerical results for a chosen benchmark point in figure~\ref{fig: numVsana_nDM}. The left panel shows the evolution of the energy densities of the SM radiation bath and the scalar field, as well as the number density of DM particles (multiplied with the DM mass to obtain the unit of an energy density), and demonstrates good agreement between the numerical and analytical curves. Furthermore, to emphasize the difference between phase-in and the standard freeze-in scenario, we show also the case of adiabatic evolution in dashed lines. In this case, we ignore the entropy injection from the scalar field decay, such that $\rho_\mathrm{SM}\propto a^{-4}$ during all stages and $n_\mathrm{DM}\propto a^{-3}$ after the production of the initial abundance during stage I. The contributions to DM production in each stage are detailed in the right panel of figure~\ref{fig: numVsana_nDM}.
 Both plots show that the number density of DM is enhanced because of the decay of the scalar (production in stages III and IV). The values of DM relic density obtained from the numerical and the analytical solutions are also presented in the right panel and show very good agreement. 

\begin{figure}
    \centering
    \begin{subfigure}{0.48 \textwidth}
    \centering
        \includegraphics[width=\textwidth]{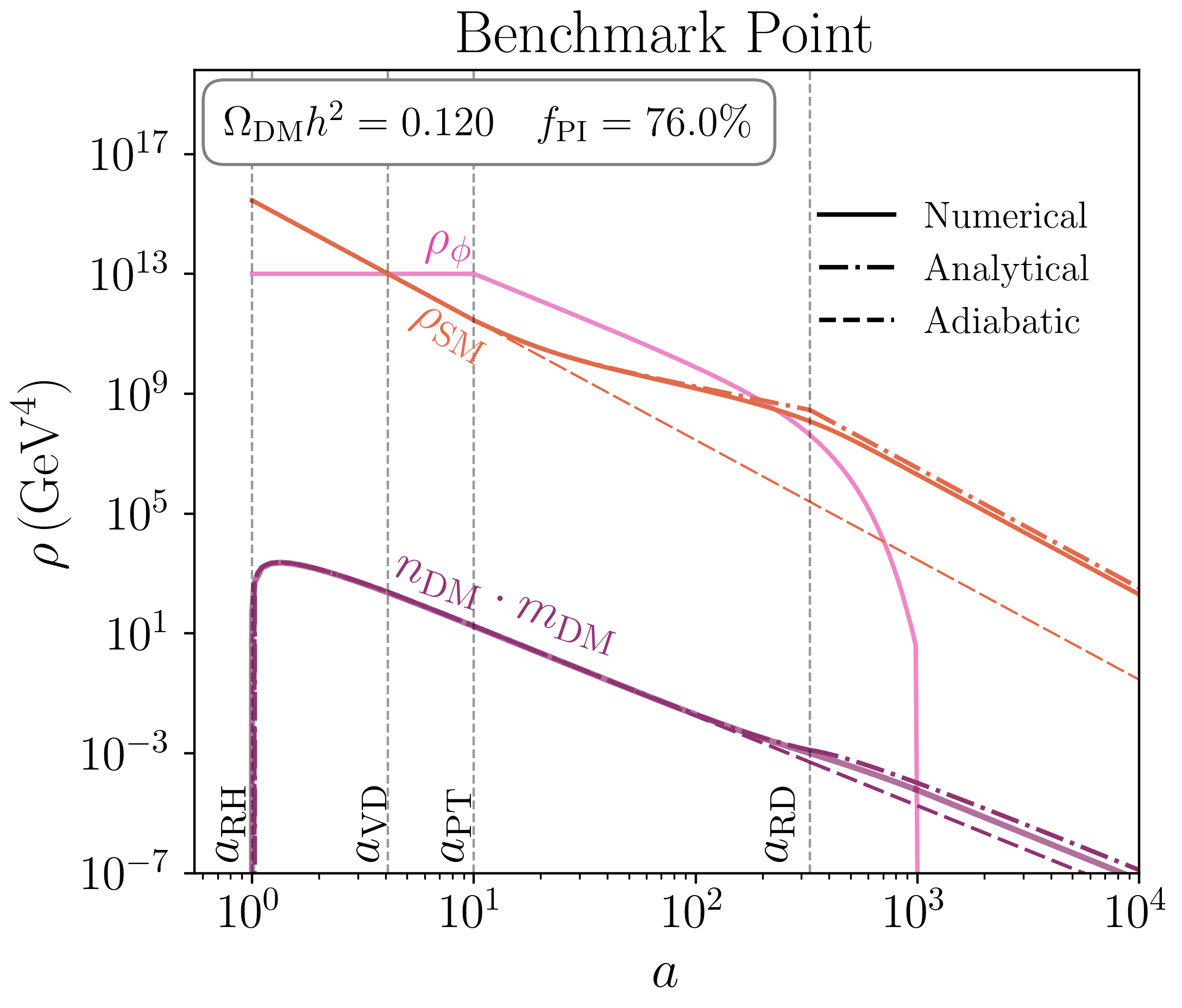}
    \end{subfigure}
    \begin{subfigure}{0.48 \textwidth}
    \centering 
        \includegraphics[width=\textwidth]{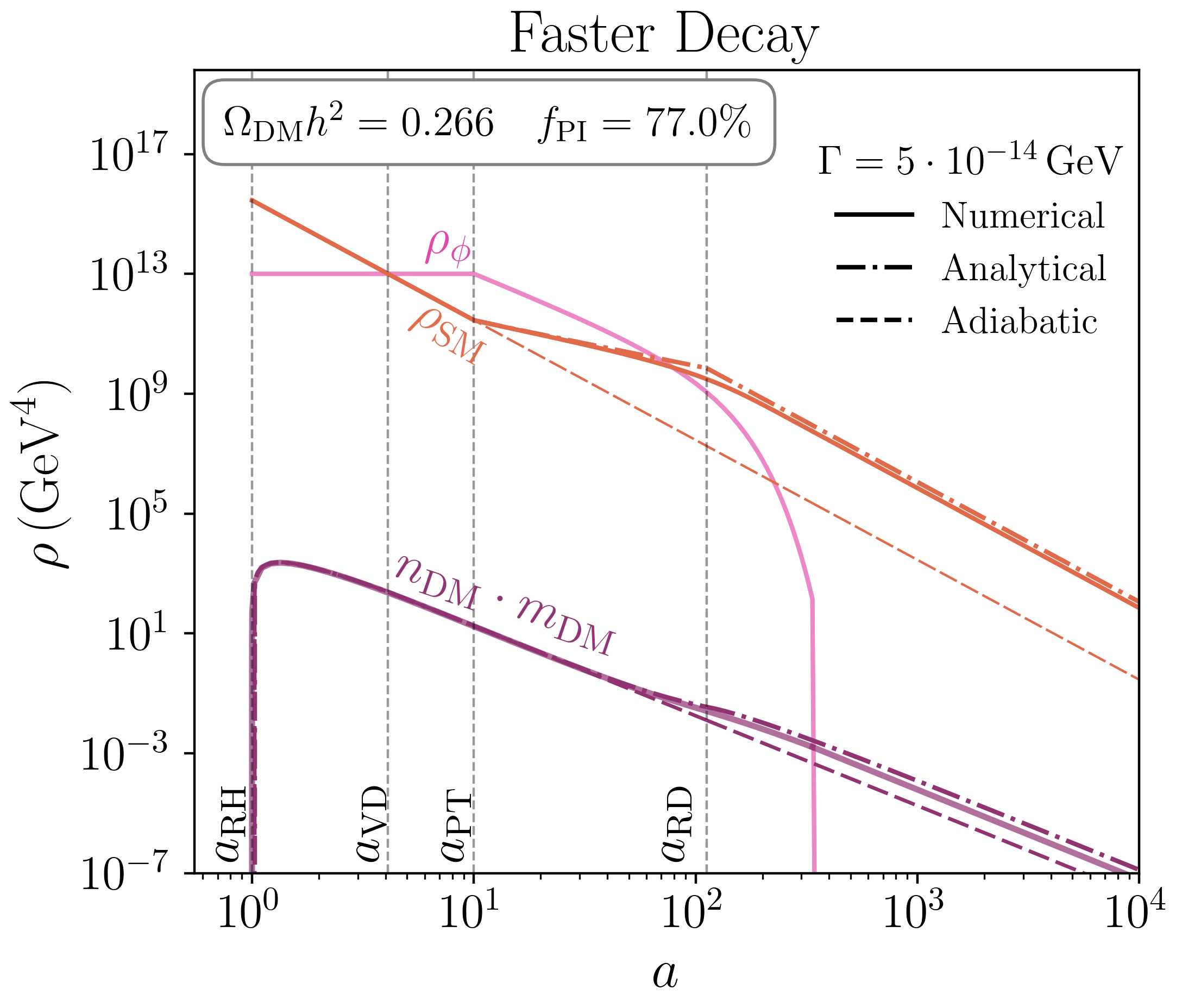}
    \end{subfigure}
    \begin{subfigure}{0.48 \textwidth}
    \centering
        \includegraphics[width=\textwidth]{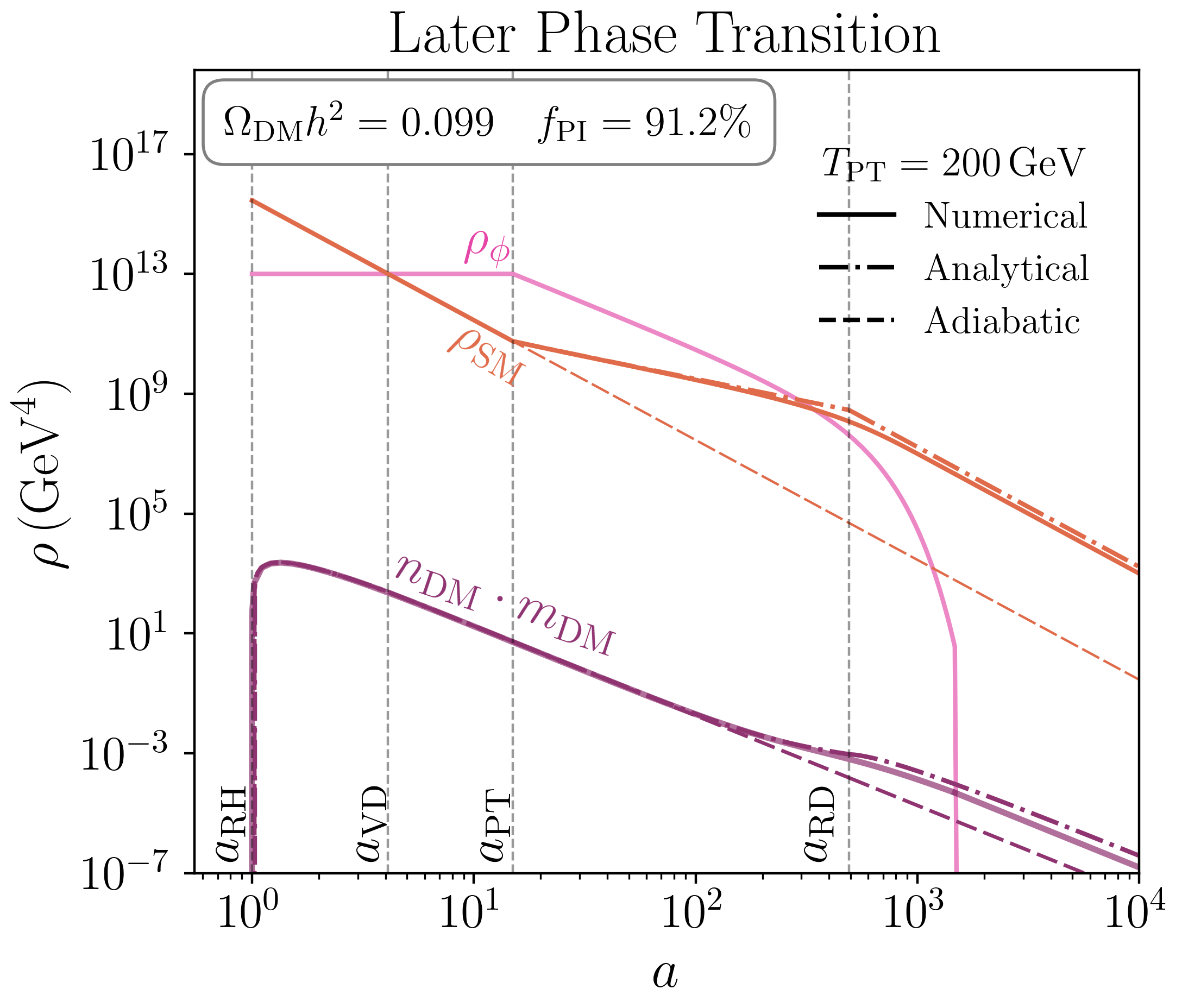}
    \end{subfigure}
    \begin{subfigure}{0.48 \textwidth}
    \centering
        \includegraphics[width=\textwidth]{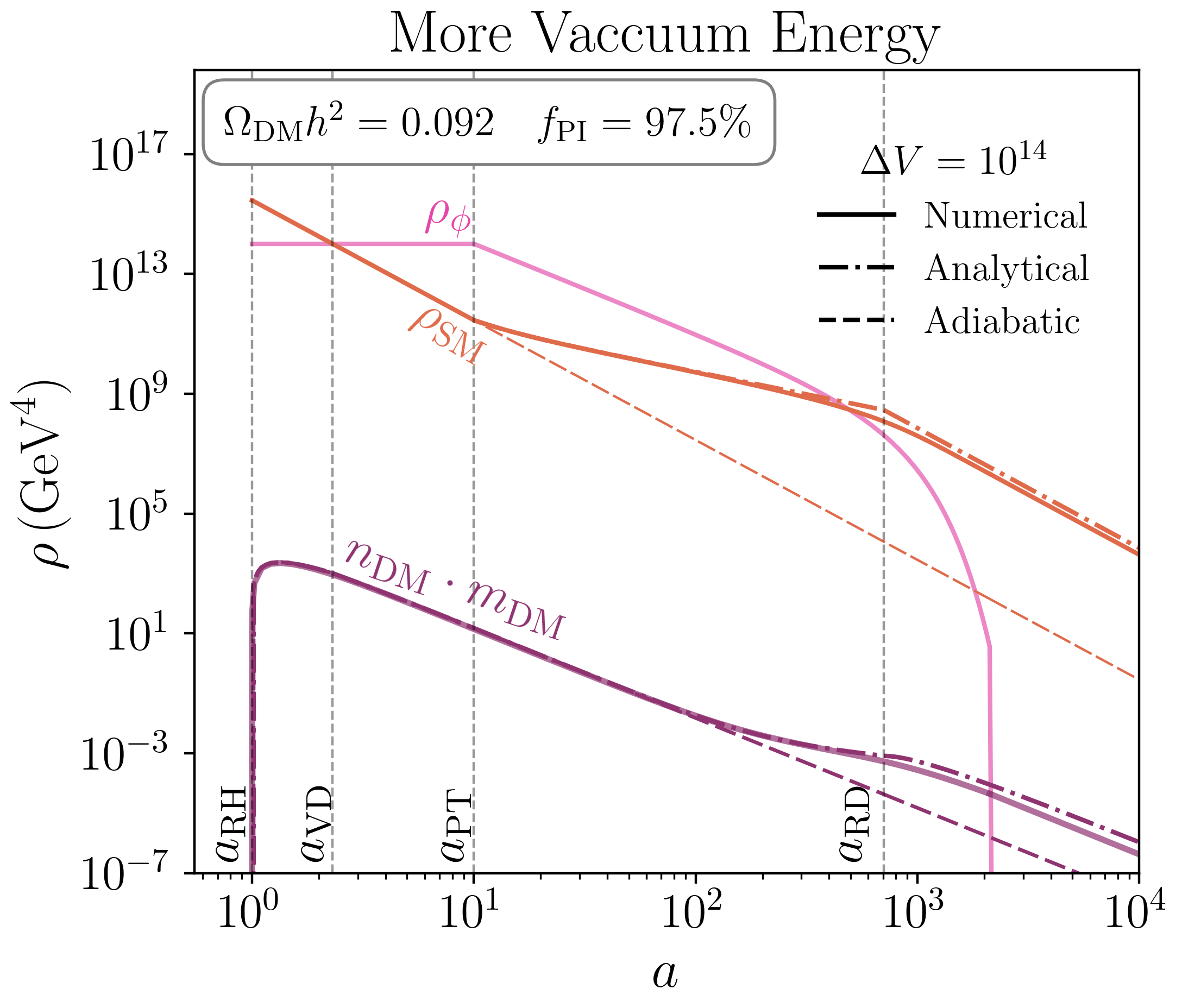}
    \end{subfigure}
    \begin{subfigure}{0.48 \textwidth}
    \centering
        \includegraphics[width=\textwidth]{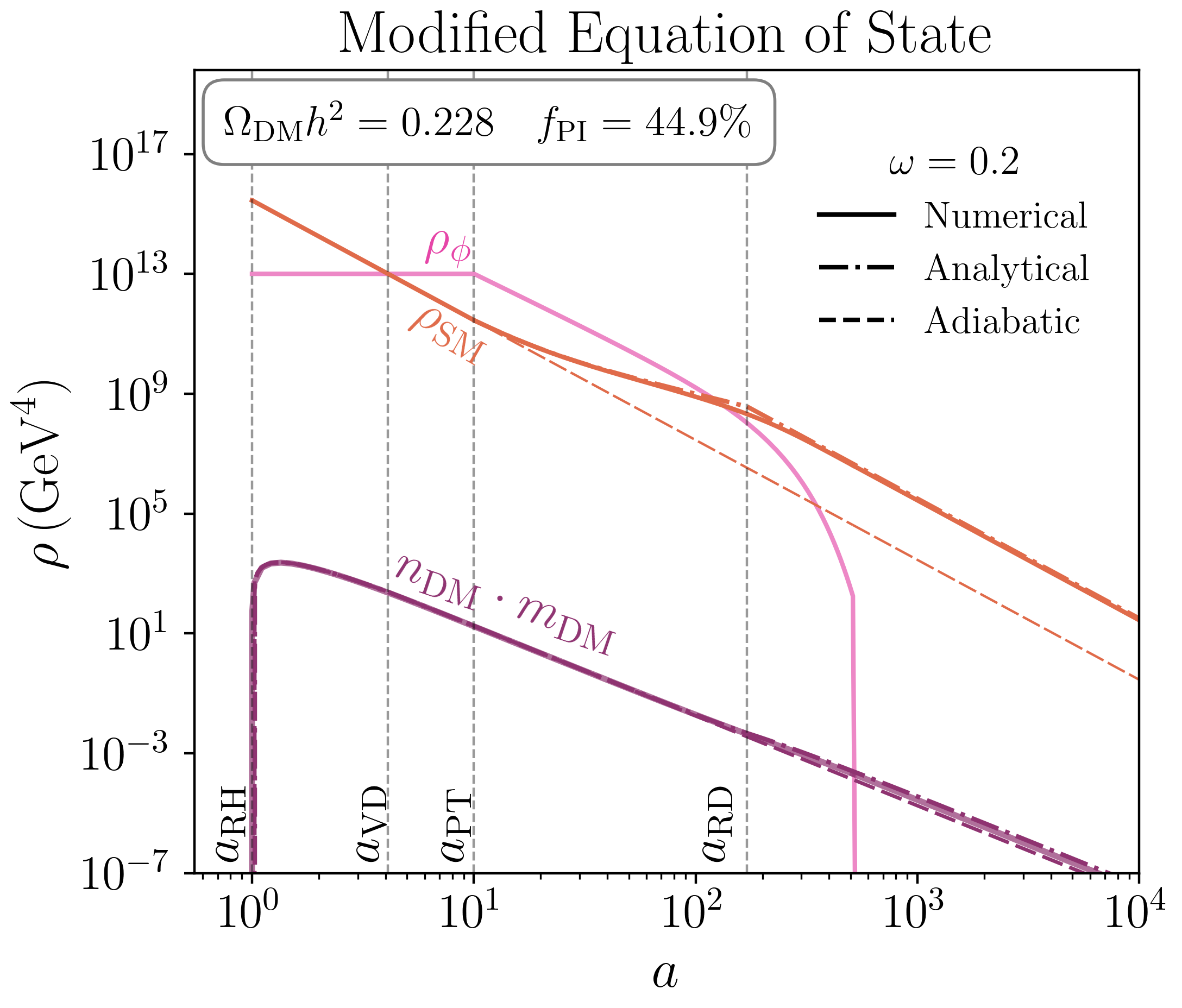}
    \end{subfigure}
    \begin{subfigure}{0.48 \textwidth}
    \centering
        \includegraphics[width=\textwidth]{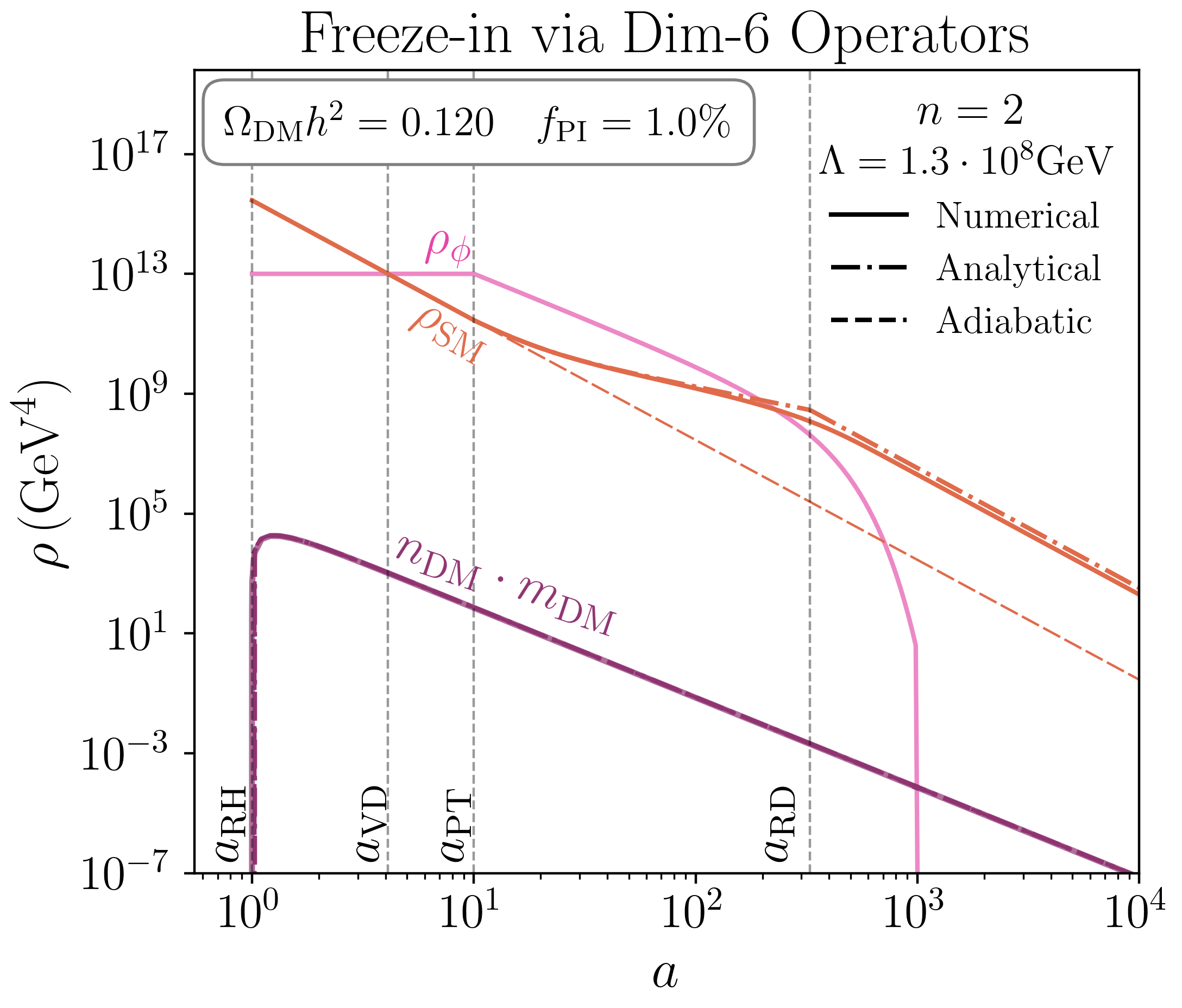}
    \end{subfigure}
    \caption{The upper left panel corresponds to the same benchmark point as in figure~\ref{fig: numVsana_nDM}, with $\omega = 0$ and $n = 1$. In each of the other panels we show the impact of varying individual parameters. The value $f_\mathrm{PI}$ represents the fraction of DM relic density produced through phase-in. Note that in the bottom-right panel both the value of $n$ and the value of $\Lambda$ are modified, with the latter being fixed by the relic density requirement. The benchmark values are: $m_\mathrm{DM}=1\,\mathrm{MeV}$, $T_\mathrm{RH}=3\cdot10^3\,\mathrm{GeV}$, $T_\mathrm{PT}=300\,\mathrm{GeV}$, $\Delta V = 10^{13}\,\mathrm{GeV}^4$, $\Gamma = 10^{-14}\,\mathrm{GeV}$ and $\Lambda=1.88 \cdot 10^{13}\,\mathrm{GeV}$.}
    \label{fig: varying parameters}
    \vspace{-1 cm}
\end{figure}  

The DM abundance and whether it is dominated by freeze-in or phase-in depends on the interplay of multiple parameters: the temperature after inflationary reheating $T_\mathrm{RH}$, the properties of the scalar field and its potential -- characterized by $T_\mathrm{PT}$, $\Delta V$, $\Gamma$ and $\omega$ -- as well as the dimension of the non-renormalizable interaction between SM and DM, encoded in the parameter $n$. In figure~\ref{fig: varying parameters}, we show the effect of varying individual parameters starting from the same benchmark point as in figure~\ref{fig: numVsana_nDM}. For this set of figures $T_\mathrm{RH}$ is kept constant. In each case, the total abundance of DM relics, $\Omega_\mathrm{DM}h^2$, and the fraction of DM originating from phase-in, $f_\mathrm{PI}$, are calculated. The latter is defined as: 
\begin{equation}
    f_\mathrm{PI} = \frac{\Omega_\mathrm{DM,PI} h^2}{\Omega_\mathrm{DM} h^2},  
    \label{eq:fPI}
\end{equation}
where $\Omega_\mathrm{DM,PI}$ is obtained by considering only the  contributions to DM  production after the phase transition (stages III and IV). 

The top-left panel of figure~\ref{fig: varying parameters} is identical to figure~\ref{fig: numVsana_nDM}. In this case, phase-in accounts for 76\% of the final DM abundance. In the top-right panel we consider the case of slower decays of the scalar field. While this increases the overall abundance of DM relative to the benchmark point, the relative contribution from phase-in remains almost constant.  Indeed, in the next section we will show that our analytical approximation predicts that for $n=1$ and $\omega=0$ the phase-in fraction is independent of $\Gamma$.

The two panels in the middle row of figure~\ref{fig: varying parameters} illustrate that more strongly supercooled phase transitions, i.e. a smaller $T_\text{PT}$ or higher values for the latent heat $\Delta V$, lead to more dilution of the initial DM abundance and an enhancement of the phased-in contributions. The two panels in the bottom row show that phase-in is more difficult to achieve if the equation of state of the decaying scalar field has $\omega > 0$, or if the interaction operator is of higher dimension ($n > 1$). 

Overall, we find that phase-in is most efficient when the decaying scalar field behaves like matter $\omega=0$, for lower dimensions of the interaction and for strongly supercooled phase transitions. In the following section we will quantify these findings more systematically.

\section{Conditions for dark matter phase-in}
\label{sec:phase-in}

\begin{figure}
    \centering    \includegraphics[width=0.7\linewidth]{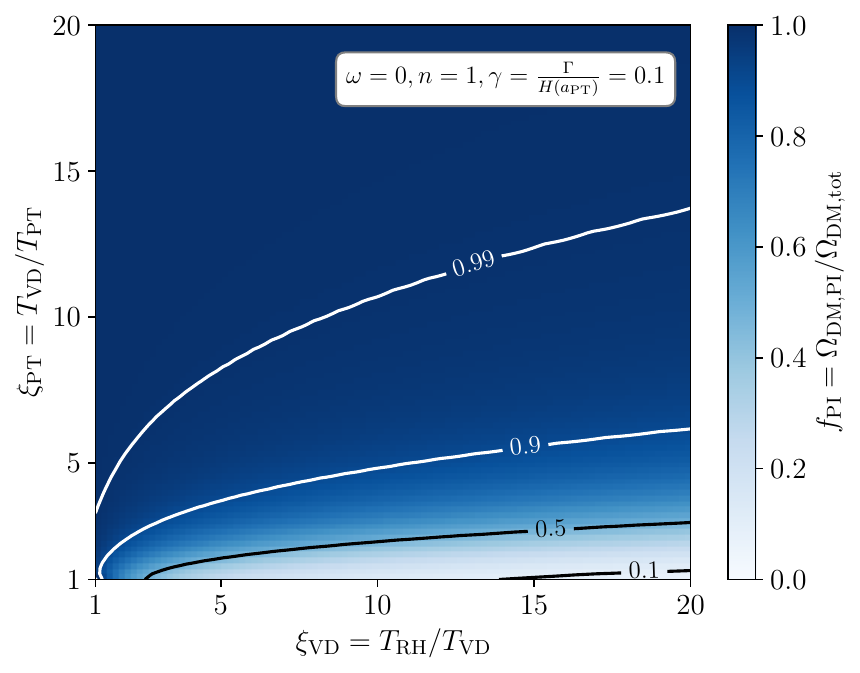}
    \caption{Fraction $f_\text{PI}$ of DM produced via phase-in as a function of the dimensionless temperature ratios $\xi_\text{VD} = \frac{T_\text{RH}}{T_\text{VD}}$ and $\xi_\text{PT} = \frac{T_\text{VD}}{T_\text{PT}}$ for interactions via a dimension-5 operator ($n=1$) and assuming that the scalar field behaves like matter after the phase transition ($\omega=0$). The dimensionless decay rate is $\gamma = \frac{\Gamma}{H(a_\text{PT})} = 0.1$, corresponding to a slow decay of the scalar field.  
    }
    \label{fig:DMfraction}
\end{figure}

In this section, we study how the fractional phase-in contribution $f_\text{PI}$ defined in eq.~\eqref{eq:fPI} depends on the various model parameters. 
While the total amount of DM produced depends on the temperature scale when freeze-in is most efficient, the phase-in fraction is almost completely independent of $T_\text{RH}$ (except for the implicit temperature dependence of the effective degrees of freedom $g_\ast$). For the purpose of this section, we will therefore fix the reheating temperature to $T_\text{RH} = 1 \, \mathrm{TeV}$.  Moreover, $f_\text{PI}$ is also independent of $\Lambda$ and $m_\text{DM}$. 

The phase-in fraction can on dimensional grounds depend on the various model parameters only through dimensionless ratios. Keeping in mind that the vacuum energy $\Delta V$ and the temperature at the beginning of vacuum domination $T_\text{VD}$ are related via $\rho_\text{SM}(T_\text{VD}) = \Delta V$, we choose the following three combinations:
\begin{align}
    \xi_\text{VD} & \equiv \frac{T_\text{RH}}{T_\text{VD}} \; ,&
    \xi_\text{PT} & \equiv \frac{T_\text{VD}}{T_\text{PT}}  \; , &
    \gamma & \equiv  \Gamma \sqrt{\frac{3 M_\text{Pl}^2}{8 \pi \Delta V}}  \; . \label{eq:ratios}
\end{align}
The temperature ratio $\xi_\text{VD}$ quantifies how long the universe spends in radiation domination after inflationary reheating and before the onset of vacuum domination. The temperature ratio $\xi_\text{PT}$ characterises the amount of supercooling of the phase transition, i.e.\ the duration of vacuum domination. During this period of accelerated expansion, the scale factor grows approximately by $\log \xi_\text{PT}$ e-folds. The dimensionless decay constant $\gamma$ determines how quickly the universe returns to radiation domination after the phase transition. Although in our setup $\xi_\text{VD,PT} \geq 1$, the dimensionless decay constant can take values both greater than unity (corresponding to instantaneous reheating) or smaller than unity (slow reheating). 

In figure~\ref{fig:DMfraction}, we show the dependence of $f_\text{PI}$ on $\xi_\text{VD}$ and $\xi_\text{PT}$ by numerically solving the system of Boltzmann equations ({\ref{eq:Boltzmann_Eq_phi})--(\ref{eq:Boltzmann_Eq_DM}) . We consider a case in which the scalar field behaves like matter after the phase transition ($\omega = 0$), the DM production proceeds via a 5-dimensional operator ($n = 1$), and the scalar decay rate is smaller than the Hubble expansion rate at the moment of the phase transition ($\gamma = 0.1$, slow reheating).
We observe that more supercooling (larger values of $\xi_\text{PT}$) correspond to a larger fraction of the phase-in fraction, while a delayed onset of vacuum domination (larger values of $\xi_\text{VD})$ reduces the phase-in contribution. 

We can understand these findings using the analytical estimates obtained in section~\ref{sec:evolution}. As shown in eq.~\eqref{eq:DMtot}, the total DM abundance is determined by two separate contributions: the freeze-in contribution produced immediately after inflationary reheating (stage I) and phase-in contribution produced at the end of the second reheating period (stage III and IV). The former is enhanced by the larger temperature but suppressed by the dilution factor $D$ defined in eq.~\eqref{eq:dilution}, so it is a priori unclear which contribution dominates. 

Let us define the ratio of the two contributions
\begin{equation}
    r = \frac{\Omega_\text{DM,PI}}{\Omega_\text{DM,FI}}
\end{equation}
such that $f_\text{PI} = r / (1+r)$. The requirement that the DM abundance is dominated by the phase-in contribution ($f_\text{PI} > 0.5$) then translates to $r > 1$.
From eq.~\eqref{eq:DMtot}, we infer
\begin{equation}
    r = D (1 + \kappa) \left(\frac{T_\text{RD}}{T_\text{RH}}\right)^{2n-1}
\end{equation}
for the case of slow reheating, i.e.\ $\Gamma < H(a_\text{PT})$. In terms of the dimensionful quantities (i.e.\ neglecting dimensionless numerical factors of order unity) we obtain
\begin{equation}
    r \approx T_\text{RH}^{1 - 2 n} \, T_\text{PT}^{-3} \, \left(M_\text{Pl} \Gamma\right)^{1 + n - \tfrac{2}{1 + \omega}} \, \Delta V^{\tfrac{1}{1 + \omega}} g_\star^{-(n+1)/2} \; .
    \label{eq:phase-in}
\end{equation}
Clearly, the condition $r > 1$ becomes easier to satisfy if the phase transition is more strongly supercooled, i.e.\ if $T_\text{PT}$ is as small and $\Delta V$ as large as possible. For constant supercooling, $\xi_\text{PT} = \text{const}$, the phase transition should happen as early as possible, i.e.\ shortly after inflationary reheating. 

A surprising implication of eq.~\eqref{eq:phase-in} is that for $n > 1$ or $\omega > 0$ the stage-I contribution is suppressed by making $\Gamma$ as large as possible, i.e.\ by reheating the SM thermal bath as quickly as possible after the phase transition. Extending the period of scalar field domination (stage III) actually suppresses the contribution from stages III and IV relative to the one from stage I. For $n = 1$ and $\omega = 0$, eq.~\eqref{eq:phase-in} becomes independent of $\Gamma$, i.e.\ the length of stage III is inconsequential for the relative importance of the two contributions.

This conclusion holds in particular for the case of an early period of matter domination, which is recovered if we set $\omega = 0$ and assume that there is no period of vacuum domination, i.e. that $a_\text{VD} = a_\text{PT}$ or $\Delta V = \pi^2 g_\star T_\text{PT}^4 / 30$. 
In this case, we obtain
\begin{equation}
    r^\text{EMD} = T_\text{PT} T_\text{RH}^{1 - 2 n} \left(M_\text{Pl} \Gamma\right)^{-1 + n} g_\star^{-(n-1)/2} \; ,
\end{equation}
such that $r^\text{EMD} > 1$ can never be satisfied for $n \geq 1$ since $T_\text{RH} > T_\text{PT} \gtrsim \sqrt{M_\text{Pl} \Gamma}$. We conclude that for freeze-in production of DM via non-renormalisable interactions, a period of early matter domination is insufficient to suppress the freeze-in contribution relative to the phase-in contribution. In other words, while such a period does dilute the pre-existing abundance, it also suppresses phase-in production by the same amount. To enhance the latter contribution relative to the former, it is necessary to have a supercooled phase transition with $\Delta V > \pi^2 g_\star T_\text{PT}^4 / 30$.

For the case of instantaneous reheating, i.e.\ $\Gamma > H(a_\text{PT})$, we find
\begin{equation}
    r^\text{inst} = D^\text{inst} \left(\frac{T_\text{RD}}{T_\text{RH}^\text{inst}}\right)^{2n-1} \; ,
\end{equation}
which corresponds to
\begin{equation}
r^\text{inst} \approx T_\text{RH}^{-2 n + 1} T_\text{PT}^{-3} \Delta V^{\tfrac{n + 1}{2}} g_\star^{-(n+1)/2} \left(\tfrac{8\pi}{3}\right)^{\tfrac{1 + n}{2} - \tfrac{1}{1 + \omega}} \; .
\label{eq:inst_phase-in}
\end{equation}
As expected, this expression is always independent of $\Gamma$ and $\omega$ and agrees with eq.~\eqref{eq:phase-in} for $n = 1$, $\omega = 0$. The final numerical factor has been included to ensure that eqs.~\eqref{eq:phase-in} and~\eqref{eq:inst_phase-in} match for $\Gamma = H(a_\text{PT})$. In fact, we can combine the two inequalities in a more economical form and smoothly interpolate between the late reheating and instantaneous reheating case by writing
\begin{equation}
r^\text{comb} \approx  T_\text{RH}^{-2 n + 1} T_\text{PT}^{-3} \Delta V^{\tfrac{n + 1}{2}} g_\star^{-(n+1)/2} \left( 
   \tfrac{\sqrt{\Delta V}}{M_\text{Pl} \Gamma} + 
    \sqrt{\tfrac{3}{8 \pi}}\right)^{\tfrac{2}{1 + w} - 1 - n} \; ,
    \label{eq:rcomb_T}
\end{equation}
which reproduces eqs.~\eqref{eq:phase-in} and~\eqref{eq:inst_phase-in} in the limit $\gamma \ll 1$ and $\gamma \gg 1$, respectively. In terms of the dimensionless ratios defined in eq.~\eqref{eq:ratios} we find
\begin{equation}
r^\text{comb} \approx \xi_\text{PT}^3 \,  \xi_\text{VD}^{1 - 2 n}
\left(\frac{\pi^2}{30}\right)^{\tfrac{n+1}{2}} 
\left[ \sqrt{\frac{3}{8 \pi}} \left(1 + \frac{1}{\gamma}\right)\right]^{-n - 1 + \tfrac{2}{1 + \omega}} \; .
\label{eq:rcomb}
\end{equation}

\begin{figure}
    \centering
   \includegraphics[width=1.0\linewidth]{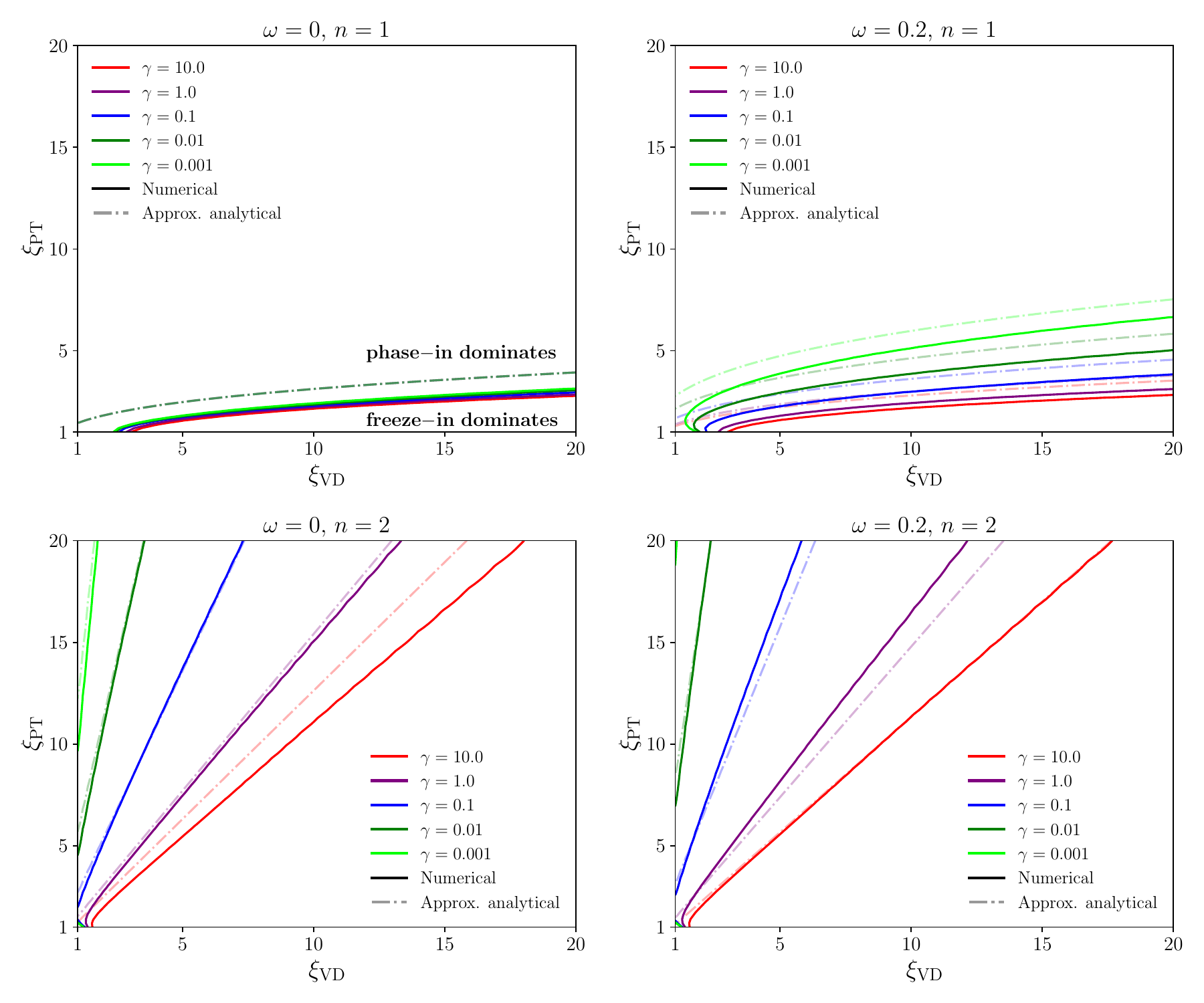}
    \caption{Illustration of the phase-in condition for several scenarios described by different combinations of values of $\omega$ and $n$ in each panel, and for different values of $\gamma = \frac{\Gamma}{H(a_\text{PT})}$, associated to different colors. 
    The phase-in condition corresponds to a phase-in fraction $f_\text{PI} = 0.5$ and marks the boundary between the region of parameter space where the DM production is dominated by the freeze-in contribution coming from stages I and II (below and to the right of each line) and the region of the parameter space where the phase-in contribution produced in stages III and IV dominates (above and to the left of each line). 
    Solid line corresponds to the phase-in condition extracted from the numerical solution of the Boltzmann equations; the dash-dotted line corresponds to the condition analytically estimated in eq. \eqref{eq:rcomb}).}
    \label{fig:phase-in_subplots}
\end{figure}

In figure~\ref{fig:phase-in_subplots}, we show the values of $\xi_\mathrm{VD}$ and $\xi_\mathrm{PT}$ that correspond to the boundary between freeze-in and phase-in, i.e.\ $f_\mathrm{PI}=0.5$. The parameter region above and to the left of a given curve corresponds to DM production dominated by phase-in, while freeze-in provides the main contribution below and to the right. In the two columns, we consider different equations of state for the scalar field: a matter-like behaviour $(\omega=0)$ and an intermediate behavior between radiation and matter $(\omega=0.2)$. In the two rows, we consider DM production via 5- and 6-dimensional operators corresponding respectively to $n=1$ and $n=2$. Curves of different colors correspond to different values of the dimensionless decay rate $\gamma$ defined in eq.~\eqref{eq:ratios}. Finally, we compare the analytical approximation in eq.~\eqref{eq:rcomb},  represented by dash-dotted lines, with the detailed numerical solutions of the Boltzmann equations represented by solid lines.

In the top-left panel ($\omega = 0$ and $n=1$), all the analytical curves coincide. The numerical results, on the other hand, show a mild dependence on the decay rate, which indicates that the cancellation of $\gamma$ (or equivalently of $\Gamma$) in the analytical expressions is due to the approximations made in the derivation, see the discussion following eq.~\eqref{eq:phase-in}. Nonetheless, the results show that phase-in is dominant in nearly all of the parameter space considered. 
In particular, even for large values of $\xi_\text{VD}$ up to $20$, phase-in constitutes the main contribution to the DM abundance, provided that $\xi_\text{PT} \gtrsim 3$, i.e.\ that the phase transition is preceded by more than one e-fold of vacuum domination.

In the top-right panel ($\omega = 0.2$ and $n=1$), we observe a more pronounced dependence of the phase-in condition on the decay rate $\gamma$. This is because for $\omega > 0$, the energy density of the scalar field decreases more rapidly than for $\omega=0$, such that the dilution effect (and hence the relevance of the phase-in contribution) is suppressed  for slower decays (smaller $\gamma$). 
For the case of instantaneous reheating ($\gamma = 10$), the result does not depend on $\omega$ and therefore agrees with the corresponding curve in the top-left panel. 
We also note that for small values of $\xi_\text{VD}$, the analytical curves begin to deviate from the numerical results. 
This difference can be explained by the fact that the simplifying assumptions used in the derivation break when stages I and II becomes too short.

The two panels in the bottom row, where $n=2$, show that a dominant phase-in contribution is more difficult to achieve if the cross section of DM production depends more strongly on the temperature of the SM bath. 
Compared to the case with $n=1$, much larger values of $\xi_\text{PT}$, i.e.\ much stronger supercooling, are needed for the phase-in condition to be satisfied. 
The intuitive explanation is that the stronger temperature dependence enhances the freeze-in contribution compared to that of phase-in, since the former takes places at much higher temperatures.
This enhancement becomes larger with increasing $\xi_\text{VD}$ but can be compensated by a larger amount of dilution, corresponding to larger $\xi_\text{PT}$. 
Moreover, to avoid a further decrease of the temperature of the SM thermal bath after the phase transition, the scalar field should decay as quickly as possible after the phase transition, in particular for $\omega = 0.2$, as considered in bottom-right panel. 
Nevertheless, even in this case, phase-in can be dominant for sufficiently large values of $\xi_\text{PT}$ and $\gamma$ and sufficiently small values of $\xi_\text{VD}$.

In general, figure~\ref{fig:phase-in_subplots} confirms our conclusion from section~\ref{sec:comparison}, namely that the full numerical results can be quite accurately reproduced with our analytical approximations. These findings justify \emph{a posteriori} the various simplifications and imply that the analytical condition in eq.~\eqref{eq:rcomb} provides a useful guidance to estimate the relevance of the phase-in contribution and to determine the most interesting regions of parameter space. Nevertheless, the agreement is clearly not perfect, highlighting the importance of solving the system of Boltzmann equations numerically when a higher level of precision is needed.

\section{Phenomenological implications}
\label{sec:pheno}

\subsection{Decaying axion-like particles and nano-Hertz gravitational waves}

As we have seen in the previous section, dark matter phase-in requires a strongly supercooled phase transition and prefers a quick reheating of the thermal bath after the end of the phase transition. As a result, the dark matter abundance becomes directly sensitive to the temperature $T_\text{RD}$, that encodes the decay rate of the scalar $\Gamma$, or, in the instantaneous reheating limit, its energy density before the phase transition $\Delta V$. Since a strongly-cooled first-order phase transition is expected to give rise to a strong gravitational wave background, this temperature can potentially be measured. Indeed, the peak frequency $f_\text{peak}$ of the stochastic gravitational wave spectrum is expected to be directly proportional to the temperature of the SM thermal bath after the phase transition~\cite{Bringmann:2023iuz}:
\begin{equation}
    f_\text{peak} = 1 \, \mathrm{mHz} \left(\frac{T_\text{RD}}{1 \, \mathrm{TeV}}\right) \left(\frac{\beta/H}{10}\right) D^{-1/3} \; ,
\end{equation}
where $\beta/H$ characterises the speed of the phase transition and a value $\beta / H \sim 10$ is typical for phase transitions with strong supercooling and large gravitational wave production.

Let us for concreteness consider the case of an axion-like particle (ALP) coupled to photons via the interaction term
\begin{equation}
    \mathcal{L}_\text{int} = \frac{g_{a\gamma}}{4} a F^{\mu\nu} F_{\mu\nu} \; .
\end{equation}
Since the effective coupling $g_{a\gamma}$ has mass dimension $-1$, the interaction is non-renormalisable, and freeze-in production of ALPs is UV dominated. Indeed, it was shown in ref.~\cite{Bolz:2000fu} that the cross section for ALP production via the Primakoff process is independent of temperature, corresponding to eq.~\eqref{eq:DM_production} with $n = 1$.\footnote{See refs.~\cite{Jain:2024dtw,Becker:2025yvb} for recent refinements in the calculation of the freeze-in production of ALPs. } For ordinary freeze-in, the resulting would-be abundance of ALPs is found to be~\cite{Balazs:2022tjl}
\begin{equation}
    \tilde{\Omega}_a h^2 = 5 \times 10^{-7} \left(\frac{m_a}{1 \, \mathrm{MeV}} \right) \left(\frac{T_\text{RH}}{10 \, \mathrm{MeV}}\right) \left(\frac{g_{a\gamma}}{10^{-12} \, \mathrm{GeV}^{-1}}\right)^2 \; .
    \label{eq:ALP-fi}
\end{equation}
However, for the couplings and masses chosen in the estimate above, the ALPs actually do not survive until the present day, instead, they decay with a lifetime of approximately
\begin{equation}
    \tau_a \approx 10^{11} \, \mathrm{s} \left(\frac{g_{a\gamma}}{10^{-12} \, \mathrm{GeV}^{-1}}\right)^{-2} \left(\frac{m_a}{1 \, \mathrm{MeV}} \right)^{-3} \; .
\end{equation}

The example values of $m_a$, $T_\text{RH}$ and $g_{a\gamma}$ in the above equations have been chosen to satisfy all current experimental constraints~\cite{Balazs:2022tjl}. However, the effect of ALPs with such masses and couplings on the CMB spectral shape can potentially be observed by the proposed PIXIE mission~\cite{Kogut:2024vbi}. Measuring both $\mu$ and $y$ distortions would make it possible to infer both $\tau_a$ and $\tilde{\Omega}_a h^2$~\cite{Cadamuro:2011fd}, so that $m_a$ and $g_{a\gamma}$ can be inferred if $T_\text{RH}$ is known.

Determining $T_\text{RH}$ from cosmological data is generally challenging. However, if ALPs are produced predominantly through the phase-in mechanism, $T_\text{RH}$ in eq.~\eqref{eq:ALP-fi} is replaced by the temperature $T_\text{RD}$ after the phase transition. This temperature may be directly measurable if the phase transition at the same time generates an observable gravitational wave signal. In the example above, the interesting temperature range is $T_\text{RD} \sim 10 \text{--}100 \, \text{MeV}$, corresponding to gravitational wave frequencies $f_\text{peak} \sim 10 \text{--} 100 \, \mathrm{nHz}$, i.e.\ in the range probed by pulsar timing arrays. A combination of these measurements with future measurements of CMB spectral distortions may hence reveal ALPs produced via the phase-in mechanism.

\subsection{Sterile neutrinos as a mixture of warm and cold dark matter}

So far, we have only discussed the DM number density, assuming that the kinetic energy redshifts fast enough to be irrelevant for structure formation. For ordinary freeze-in production of DM, this is typically a good approximation for $m_\text{DM} \gtrsim 15 \, \mathrm{keV}$~\cite{Decant:2021mhj}. For smaller masses, the free-streaming of DM particles can have an observable effect on small-scale structure formation, which can be probed in particular with measurements of the Lyman-$\alpha$ forest.

Since DM particles produced via freeze-in do not in general follow a thermal distribution~\cite{Murgia:2017lwo}, a detailed study of these constraints requires solving the Boltzmann equation at the phase-space level in order to predict the linear matter power spectrum $\mathcal{P}(k)$, which requires model-specific calculations of the relevant cross sections. Nevertheless, for a first estimate, we can assume that DM particles produced directly from the thermal bath at temperatures much larger than the DM mass will inherit a kinetic energy comparable to the temperature of the thermal bath, such  that the phase space distribution will approximately resemble a thermal distribution with temperature comparable to the one of the SM thermal bath.

In the case of phase-in production of DM, however, the situation becomes more interesting. The DM particles produced before the phase transition (at temperatures close to $T_\text{RH}$) will initially have similar kinetic energies as the SM bath particles. However, the entropy injected into the SM thermal bath after the phase transition increases the temperature of the latter relative to the former. Once the universe has returned to radiation domination after the phase transition, the typical kinetic energy of the DM particles will be of the order of $T / D^{1/3}$ with the dilution factor $D$ defined in eq.~\eqref{eq:dilution}. As a result, the Lyman-$\alpha$ forest bound on the warm dark matter mass is relaxed by a factor $D^{1/3}$, and it becomes possible to have cold dark matter even for sub-keV masses.

The phase-in contribution, on the other hand, will yield DM particles with much larger kinetic energy, comparable to $T$, such that the phase space distribution becomes bimodal. Effectively, dark matter behaves like a mixture of warm and cold dark matter, even though it comprises a single particle species, because production happens at two different points in the evolution of the universe. Such a mixture leads to a characteristic step-like feature in the transfer function $\mathcal{T}(k) = (\mathcal{P}(k) / \mathcal{P}_\text{cdm}(k))^{1/2}$, where $\mathcal{P}_\text{cdm}$ denotes the linear matter power spectrum of cold dark matter. The position of the step is determined by the warm dark matter mass, while the height of the step is determined by the fraction of warm dark matter $f_\text{wdm}$~\cite{Boyarsky:2008xj}.

Let us for concreteness consider a sterile neutrino $N$ produced from interactions of SM fermions $f$ via a vector mediator that is heavy compared to the reheating temperature $T_\text{RH}$. The effective interaction can then be written as
\begin{equation}
    \mathcal{L}_\text{int} = \frac{1}{\Lambda^2} \bar{N} \gamma^\mu N \bar{f} \gamma_\mu f \; ,
\end{equation}
corresponding to UV-dominated freeze-in with $n = 2$. Setting for example $m_N = 1 \, \mathrm{keV}$, the phase-in contribution would correspond to warm dark matter. For $D \gtrsim 5000$, the freeze-in contribution, on the other hand, is indistinguishable from cold dark matter. Such a dilution can be achieved for example for $T_\text{PT} = 200 \, \mathrm{GeV}$, $\Delta V = 10^{14} \, \mathrm{GeV}^4$, $\Gamma = 10^{-14} \, \mathrm{GeV}$ and $\omega = 0$.

With these parameters fixed, the value of $f_\text{wdm}$ depends exclusively on $T_\text{RH}$ (assuming that $\Lambda$ is adjusted in such a way that the total abundance of sterile neutrinos matches the observed value). Numerically, we find
\begin{equation}
    f_\text{wdm} \approx \frac{1}{1 + \left(\frac{T_\text{RH}}{1850 \, \mathrm{GeV}}\right)^3}
\end{equation}
To satisfy the observational upper bound $f_\text{wdm} < 0.25$ for $m_N = 1 \, \mathrm{keV}$~\cite{Hooper:2022byl}, we thus require $T_\text{RH} \gtrsim 2.7 \, \mathrm{TeV}$.

\section{Conclusions}
\label{sec:conclusions}

The freeze-in mechanism offers an attractive alternative to the WIMP paradigm by extending the relevant parameter space of dark matter (DM) models to smaller couplings. It however introduces a new complication, namely the sensitivity of the predicted DM abundance to initial conditions. This sensitivity is particularly severe in the case of UV-dominated freeze-in via non-renormalisable interactions, for which most of DM is produced at the highest temperatures of the Standard Model (SM) thermal bath. As a result, predictions depend on the reheating temperature $T_\mathrm{RH}$ as well as on the details of inflationary reheating, which are difficult to constrain observationally.  
 
In this work, we investigated the sensitivity of DM freeze-in to the dynamics of the early Universe in a general set-up that includes a cosmological first-order phase transition. We showed that in this setting DM production via non-renormalisable interactions is not always dominated by the highest temperatures of the SM thermal bath, but instead may be governed by the period immediately after the phase transition, during which the scalar field transfers its energy density to the SM thermal bath. We refer to this alternative production regime as \textit{DM phase-in}.

Concretely, we considered a radiation-dominated post-inflationary universe, characterized by the reheating scale $T_\mathrm{RH}$, and assumed that DM is produced from the radiation bath via a non-renormalizable operator of dimension $4+n$. As the SM temperature decreases, a scalar field trapped in a metastable vacuum becomes the largest energy component, leading to a period of vacuum domination. This sets the stage for a supercooled first-order phase transition, after which the scalar fluctuations decay and inject energy into the SM bath. If this secondary reheating dilutes the previously produced DM abundance enough, a second period of DM production can give a relevant or even leading contribution to the final DM abundance.

We derived a system of coupled Boltzmann equations that describe the evolution of the energy and number densities of the different components of the universe (SM particles, scalar field, dark radiation and DM), and solved them both numerically and (with some approximations) analytically.
We found that the final abundance of DM can be split into two separate contributions: (\textit i) the freeze-in contribution produced immediately below $T_\mathrm{RH}$, and (\textit{ii}) the phased-in contribution generated after the phase transition. The former contribution depends on the details of the phase transition and the subsequent dilution, such as the amount of vacuum energy $\Delta V$, and the equation of state $\omega$ and the decay rate $\Gamma$ of the scalar field after the phase transition. The amount of DM produced via phase-in, on the other hand, is sensitive almost exclusively to the temperature $T_\mathrm{RD}$ when the universe returns to radiation domination.

Using the approximate solutions, we derived a simple analytical condition in terms of the model parameters that can be used to determine whether DM production is dominated by freeze-in or phase-in, see eq.~\eqref{eq:rcomb_T}. We validated this condition against the full numerical solution to verify that it applies both to the case of instantaneous reheating after the phase transition and to slow decays of the scalar field. Furthermore, our findings are applicable for different equations of state of the scalar field after the phase transition.

We conclude that phase-in can be easily achieved in different scenarios even with relatively small amounts of supercooling. Our detailed results are shown in figure~\ref{fig:phase-in_subplots}.  We find that the relative phase-in contribution is enhanced for more strongly supercooled phase transitions and for earlier onsets of vacuum domination. Crucially, a period of vacuum domination is essential for the phase-in mechanism to work: an early period of matter domination followed by reheating of the SM thermal bath via scalar field decays is not enough to suppress the relative contribution of the pre-existing DM relics for any value of $n \geq 1$ (i.e.\ any type of UV-dominated freeze-in).

In fact, for $n > 1$ or $\omega > 0$, the phase-in contribution is suppressed relative to the early freeze-in production if the scalar field decays only slowly after the phase transition. In other words, the relevance of phase-in is maximized for the case of instantaneous reheating after the phase transition. For $n = 1$ and $\omega = 0$, on the other hand, the phase-in condition becomes independent of $\Gamma$, i.e.\ the duration of matter domination after the phase transition is irrelevant for the relative importance of phase-in.

Our results provide an important new perspective on the UV sensitivity of freeze-in scenarios with low reheating temperature. Provided the conditions determined in our analysis are met, the presence of a supercooled first-order phase transition makes it possible to disregard the details of inflationary reheating and calculate the final DM abundance in terms of the temperature $T_\mathrm{RD}$ of the SM thermal bath after the phase transition. Apart from its conceptual simplicity, this set-up also opens up exciting observational possibilities. If the DM relics are to a large extent generated through phase-in, the resulting DM abundance is directly related to the peak frequency of the GW signals associated with the supercooled phase transition. 

Furthermore, the phase-in scenario makes interesting predictions even in cases where its contribution does not dominate the final abundance. Since the production of DM takes place at two different times separated by a large amount of entropy injection, the final DM distribution can become a mixture of warm and cold DM. Our set-up therefore provides additional motivation to constrain the warm DM fraction $f_\mathrm{WDM}$, which is an important target for ongoing and future cosmological missions, such as DESI and EUCLID.

In the present work, we focused on the case where the mass of the DM particle is negligible compared to all relevant temperature scales, such that the DM production has a power-law dependence on the temperature of the SM thermal bath. Several recent studies explored the alternative case where the DM mass is larger than the reheating temperature, such that freeze-in production becomes Boltzmann-suppressed and depends exponentially on the SM temperature~\cite{Cosme:2023xpa,Boddy:2024vgt,Arcadi:2024wwg,Cox:2024oeb}. In this case, freeze-in may happen at substantially larger couplings and therefore within the reach of laboratory experiments. An exciting direction for future work will be to also explore ``phase-in at larger couplings'', i.e.\ to determine whether a supercooled first-order phase transition may sufficiently dilute the pre-existing DM abundance in order for the phase-in contribution to become relevant. In this case, it may be possible to correlate gravitational wave signals and cosmological data with results from laboratory experiments in order to pin down the properties of DM and constrain the evolution of the early universe.

\acknowledgments

We thank Nicolás Bernal, Yann Gouttenoire and Sebastian Hoof for discussions. This work is funded by the Deutsche Forschungsgemeinschaft (DFG) through
Grant No. 396021762 -- TRR 257. FK thanks the Instituto de Física Teórica at UAM-CSIC for hospitality during the final stages of this project.

\appendix

\section{Better estimates for the end of stage III}
\label{app:RD}
Naively, considering the condition $H\approx \Gamma$ to determine the end of stage III seems like a reasonable approximation. Although this allows for a rough estimate, it is not sufficient, as the final DM abundance is sensitive to $T_\mathrm{RD}$ in two ways: through $s_\mathrm{SM}(a_\mathrm{RD})$ and $n_\mathrm{DM}(a_\mathrm{RD})$. We find a better approximation of the numerical results for the relic density when we define $a_\mathrm{RD}$ as the moment of reestablishment of radiation domination, i.e. by setting the condition $\rho_\mathrm{SM}^\text{III}(a_\mathrm{RD})=\rho_\phi^\text{III}(a_\mathrm{RD})$. 

For this purpose, we use analytical estimates for the evolution of $\rho_\mathrm{SM}$ and $\rho_\phi$. As explained in sec \ref{subsec:evolution_III}, we neglect the backreaction of the decay on the scalar field energy density (eq. \ref{eq:phi_stageIII}) and its evolution can be approximated as
  \begin{equation}
      \rho_\phi^\text{III}(a)=\Delta V \left(\frac{a_\mathrm{PT}}{a}\right)^{3(1+\omega)}.
  \end{equation} 
At late times $a_\mathrm{PT}\ll a$, eq. \ref{eq:rhoSM_APT} becomes dominated by the second reheating : 
  \begin{equation}
      \rho_\mathrm{SM}^\text{III}(a)=\frac{\pi^2}{30}g_{*} T_\Gamma^4 \left(\frac{a_\mathrm{PT}}{a}\right)^{\frac{3}{2}(\omega+1)}.
  \end{equation}
Equating the previous two expressions gives the result we have used in our computations:
    \begin{align}
    \label{Eq: ARD Appendix}
    a_\mathrm{RD} = a_\mathrm{PT} \left(\sqrt{\frac{2 \pi}{3}}\frac{(5-3\omega) \sqrt{\Delta V}}{\Gamma_\phi M_\mathrm{pl}}\right)^{\frac{2}{3(1+\omega)}} \; ,
    \end{align}
with the corresponding temperature at RD
\begin{equation}
    \label{eq: TRD Appendix}
    T_\mathrm{RD}=\left(\frac{45 }{g_\star\pi^3}\right)^{1/4}\sqrt{\frac{\Gamma_\phi M_\mathrm{pl}}{(5-3\omega)}} \;.
\end{equation}

Although the expressions above differ from the approximated definitions in eq. (\ref{eq:aRD}) and eq. (\ref{eq:TRD}) only by $\mathcal{O}(1)$ factors, taking the simpler approximation overestimates the final relic density by around $35\%$ for the chosen benchmark point (see figure~\ref{fig: a_RD approximation}). To conclude, we note that $T_\mathrm{RD}^\mathrm{approx}/T_\mathrm{RD} = \sqrt{(5-3\omega)/2}$, i.e.\ the two results differ only by a factor $\sqrt{2}$ for $\omega=1/3$ and $\sqrt{5/2}$ for $\omega=0$. For the scale factor the corresponding ratio is given by:
\begin{equation}
    \frac{a_\mathrm{RD}^\mathrm{approx}}{a_\mathrm{RD}} = \left(\frac{2}{5-3\omega}\right)^{\frac{2}{3(1+\omega)}} = 
    \begin{cases}
        \left( \frac{2}{5}\right )^{2/3}  , & \omega=0 \\
        \frac{1}{\sqrt{2}} \, , & \omega=1/3 \; .  
    \end{cases}
\end{equation}
} 

\begin{figure}[t]
    \centering
    \includegraphics[width=0.55\linewidth]{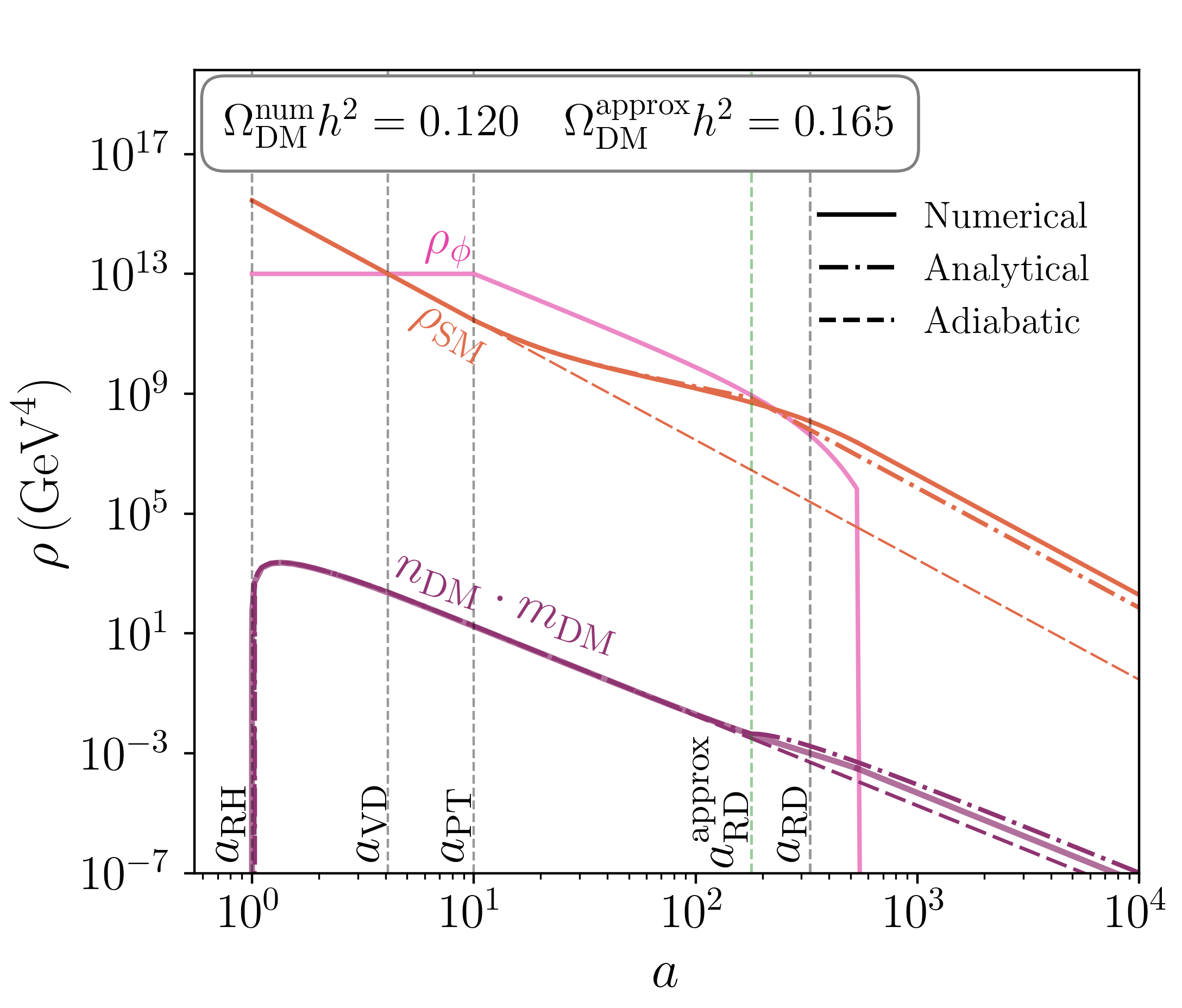}
    \caption{We show the results for the same benchmark point as in figure~\ref{fig: numVsana_nDM}, but using the approximated expressions $a_\mathrm{RD}^\mathrm{approx}$ and $T_\mathrm{RD}^\mathrm{approx}$. The green vertical line at $a_\mathrm{RD}^\mathrm{approx}$ corresponds to the condition $\Gamma \approx H$, while $a_\mathrm{RD}$ correspond to $\rho_\mathrm{SM}^\text{III}(a_\mathrm{RD})=\rho_\phi^\text{III}(a_\mathrm{RD})$. Using $a_\mathrm{RD}^\mathrm{approx}$ underestimates the SM temperature at late times, leading to a bigger relic density prediction than from the numerical solution as well as from the improved expressions (see figure ~\ref{fig: numVsana_nDM}).}
    \label{fig: a_RD approximation}
\end{figure}

\providecommand{\href}[2]{#2}\begingroup\raggedright\endgroup

\end{document}